\documentclass[11pt,titlepage,a4paper]{myarticle}

\usepackage[T1]{fontenc} 
\RequirePackage[colorlinks=true
,urlcolor=blue
,anchorcolor=blue
,citecolor=blue
,filecolor=blue
,linkcolor=blue
,menucolor=blue
,linktocpage=true
,pdfproducer=medialab
,pdfa=true
]{hyperref}
\usepackage{amsmath, mathrsfs, amsfonts, amssymb, amsthm, mathtools, graphicx, color, ucs, xparse, tikz, lmodern, physics, varioref, cleveref, tensor}

\usepackage[nosort]{cite}

\usepackage{graphicx} 
\usepackage{xcolor}
\usepackage{tikz}
\usetikzlibrary{arrows.meta,shapes.geometric}

\crefname{pluralequation}{eqs.}{eqs.}
\Crefname{pluralequation}{Eqs.}{Eqs.}
\crefformat{pluralequation}{eqs.~(#2#1#3)}
\Crefformat{pluralequation}{Eqs.~(#2#1#3)}

\renewcommand{\d}{\partial}

\newcommand{\C}{\mathbb{C}}
\newcommand{\R}{\mathbb{R}}

\newcommand{\Z}{\mathbb{Z}}

\crefname{table}{table}{tables}
\Crefname{table}{Table}{Tables}
\crefname{figure}{figure}{figures}
\Crefname{figure}{Figure}{Figures}

\newenvironment{eq}
    {\begin{equation}
    \begin{aligned}
    }
    { 
    \end{aligned}
    \end{equation}
    \ignorespacesafterend
    }

\begin{document} 

\preprint{ {\tt IFT-UAM/CSIC-26-127}\\	}

\title{\bf Massive type 0A string theory and M-theory}
\author{ Niccol\`{o} Cribiori${}^{1}$, Andriana Makridou${}^{2}$, Salvatore Raucci${}^{2,3}$
     \oneaddress{
     ${}^1$ KU Leuven, Institute for Theoretical Physics, Celestijnenlaan 200D, B-3001 Leuven, Belgium \\[.2cm]
     ${}^2$Instituto de F\'{i}sica Te\'{o}rica IFT-UAM/CSIC, C/ Nicol\'{a}s Cabrera 13--15, Campus de Cantoblanco, 28049 Madrid, Spain \\[.2cm]
     ${}^3$Departamento de F\'{i}sica Te\'{o}rica, Universidad Aut\'{o}noma de Madrid, Cantoblanco, 28049 Madrid, Spain\\ 
     {~}\\
}}

\Abstract{We propose an M-theory origin for the two massive deformations of type 0A string theory by extending Hull's realization of massive type IIA in the context of a recent proposal for massless type 0A from M-theory. We introduce an M-theory setup that we call \emph{quantum fibration}, in which two independent integral monodromies yield the Romans masses, T-dual to the two Chern numbers of the quantum fibration. A key ingredient is a formulation of the type 0B duality group in terms of a marked elliptic curve. We discuss further implications of this description and some puzzles concerning the M-theory origin of branes and Chern--Simons terms.}

\date{}
\maketitle
\setcounter{page}{1}

\tableofcontents

\section{Introduction}

Dualities have been studied for decades, yet we have likely uncovered only a small part of a much broader picture. Their best understood realizations arise in supersymmetric string compactifications, where classical geometry and topology remain applicable even at strong coupling and play a central role.

It has recently been proposed in~\cite{Baykara:2026gem}, and refined in~\cite{Altavista:2026evd,Baykara:2026vdc,Altavista:2026brr}, that dualities can be extended by introducing a notion of compactification that goes beyond conventional setups.
The conjecture of~\cite{Baykara:2026gem} is that weakly coupled type 0A string theory arises from M-theory \emph{compactified} on the singular one-dimensional space $S^1\vee S^1$, the wedge sum of two circles, which we refer to as the wedge factors. This space is assumed to be sub-Planckian, and a classical geometric description is no longer available. This is not a standard Kaluza--Klein compactification, in which all fields propagate on the same smooth internal space. Instead, the singular junction point of $S^1\vee S^1$ is treated as frozen, and the various components of the M-theory fields are assigned different resolution properties at that point; equivalently, different field components live on distinct resolutions of the singular space. Together with suitable boundary conditions, the conjecture is that this field-dependent prescription ``geometrizes'' type 0A string theory.

Having a sub-Planckian compact space is not by itself unusual. In the ordinary type IIA/M-theory duality, the radius of the M-theory circle is $R_{11}=g_s^{2/3}L_P$, where $L_P$ is the eleven-dimensional Planck length, thus implying that $R_{11}\ll L_P$ at weak string coupling, $g_s \ll 1$. In this case, the internal space is a smooth circle with no singularity at any length scale, and all fields are organized according to the same geometry. The distinctive feature of~\cite{Baykara:2026gem} is not the sub-Planckian internal space itself, but its singularity and the field-dependent prescription for resolving it. Beyond the specific case of type 0A string theory, this approach may enlarge the class of meaningful compactifications in quantum gravity. In string theory, it brings non-supersymmetric theories (see, e.g., \cite{Mourad:2017rrl,Basile:2021vxh,Angelantonj:2024tns,Raucci:2024fnp,Leone:2025mwo,Dudas:2025ubq} for recent reviews) into the web of dualities, with several consequences already explored in~\cite{Altavista:2026evd,Baykara:2026vdc,Altavista:2026brr,Dasgupta:2026maq,Basile:2026trt,Kamal:2026msr,Kan:2026sea}. Still, the proposal is presently understood mainly at the kinematical level, and it would be important to determine whether the full dynamics can be accounted for. This would likely require a deeper understanding of the microscopic degrees of freedom of M-theory. 

There is another corner of the string duality landscape perhaps not much explored: that of mass-deformed theories. For example, type IIA string theory admits a deformation controlled by the Ramond--Ramond (RR) zero-form field strength, $F_0=m$, where $m$ is known as the Romans mass. The resulting theory becomes, at low energies, the ten-dimensional massive type IIA supergravity constructed by Romans \cite{Romans:1985tz}. Other massive deformations of supergravities have been constructed over the years in various dimensions, and it remains an open problem whether all of them admit a consistent ultraviolet completion. 

For the prototype example of massive type IIA, the dual M-theory origin is subtle: Romans supergravity cannot be obtained by either a conventional circle reduction or a Scherk--Schwarz reduction of ordinary eleven-dimensional supergravity. Moreover, there is no covariant massive or cosmological deformation of eleven-dimensional supergravity~\cite{Sagnotti:1982ez,Bautier:1997yp,Deser:1997gm}.\footnote{The construction of \cite{Howe:1997qt} gives a different massive type IIA supergravity, not the Romans theory.} This is consistent with the fact that in the supergravity regime, the strong coupling of massive type IIA is obstructed~\cite{Aharony:2010af} and there is a tadpole for the worldvolume gauge field on D0-branes, the would-be momentum modes of the M-theory circle. 
Nevertheless, in~\cite{Hull:1998vy}, Hull showed that massive type IIA string theory can be obtained from M-theory in a less direct manner. First, one relates the circle reduction of massive type IIA to a Scherk--Schwarz compactification of type IIB by a mass-dependent T-duality. This construction can then be lifted to M-theory compactified on a non-trivial $T^2$-bundle over a circle, in a limit in which both the area of the torus fiber and the radius of the base circle vanish.

One can wonder whether there is a non-supersymmetric version of this procedure. The RR sector of type 0A string theory is doubled, and in particular, there are two independent zero-form field strengths, $F_0^+= m_+$, $F_0^-= m_-$, and thus two independent massive deformations \cite{Meessen:2001wk}; we will refer to $m_+$, $m_- \in \mathbb{Z}$ as Romans masses for simplicity.\footnote{We work in units $L_S = 1$, with $L_S$ the string length. We also use $L_S= \sqrt{\alpha'}=1/M_S$.} To the best of our knowledge, no M-theory origin of these deformations has been proposed. In this work, we argue for such an origin by considering M-theory on $S^1\vee S^1$.

Without attempting to derive it from a microscopic formulation of M-theory, we take the conjecture of~\cite{Baykara:2026gem} as our starting point and show how it can accommodate the two massive deformations of type 0A string theory. We find that these arise in an appropriate zero-volume limit of M-theory on what we call a \emph{quantum fibration}. The fiber consists of two tori, one for each circle in $S^1\vee S^1$, carrying independent integral parabolic monodromies. This fibration is not an ordinary topological bundle, but must be understood in the generalized sense of~\cite{Baykara:2026gem}, with the junction point playing a central role. Our quantum-geometric interpretation relies on the Chern numbers of the two wedge factors becoming the two Romans masses after T-duality. To show this, we pass through type 0B string theory and describe its conjectured duality group using an elliptic curve with a marked point. On the locus invariant under the type 0 quantum symmetry, our construction reduces, as expected, to that of~\cite{Hull:1998vy} for massive type IIA. This marked elliptic curve description leads us to speculate on a possible extension of the usual notion of string duality beyond strictly geometric setups, which could, for example, connect type 0B to type IIB.

We also comment on charged branes and Chern--Simons couplings in type 0A string theory, revealing an apparent puzzle for which we propose a possible explanation. The crossed electric-magnetic duality relations characteristic of type 0 theories associate electric objects of one wedge factor with magnetic objects of the other, and the mixed Chern--Simons couplings exhibit a similar exchange. This suggests that a complete M-theory description must include additional data relating the two wedge factors through the junction point.

\section{Massive type IIA from M-theory}\label{sec:Hull}

M-theory compactified on a circle of vanishing size is dual to type IIA string theory in the field theory (supergravity) limit and, conjecturally, also in the full quantum theory \cite{Townsend:1995kk,Witten:1995ex}, with the identification to be understood at the level of the partition functions~\cite{Diaconescu:2000wy}.
On the other hand, massive type IIA supergravity cannot be obtained from eleven dimensions via circle reduction \cite{Sagnotti:1982ez,Bautier:1997yp,Deser:1997gm}.\footnote{One can circumvent this obstruction by introducing a modified eleven-dimensional construction equipped with a preferred Killing vector \cite{Bergshoeff:1997ak}. In appropriate coordinates, all fields are required to be independent of the isometry direction, and dimensional reduction along it reproduces Romans supergravity. However, this is not a proper eleven-dimensional theory; rather, it is an eleven-dimensional rewriting of (the bosonic sector of) massive type IIA supergravity.} This might naively suggest that the massive type IIA theory may be somewhat exotic; however, it is not very different from other string and supergravity theories.

Massive type IIA supergravity provides a simple example of a supersymmetric deformation: the Romans mass modifies the supersymmetry transformations of the fermions through non-vanishing shift terms and generates a scalar potential; similar deformations can be found, e.g., in \cite[Table 4.1]{Lust:2017aqj} for maximal supersymmetry and in \cite{Farakos:2017jme} for the minimal theory in four dimensions.
From the string theory perspective, the Romans mass corresponds to a top form RR field strength, or, equivalently, to its dual $F_0$. This is arguably no more exotic than the other RR fluxes, and T-duality relates them to one another. 
It is thus natural to expect that massive type IIA string theory is also part of the web of string dualities. In fact, a realization of this expectation was proposed by Hull in \cite{Hull:1998vy}, and we review it below.

The proposal of \cite{Hull:1998vy} builds on three observations: first, Scherk--Schwarz reductions \cite{Scherk:1979zr} give rise to mass parameters; second, type IIB string theory is dual to M-theory on $T^2$ in the limit of vanishing volume and fixed complex structure; third, massive type IIA string theory compactified on a circle of radius $R$ is T-dual to a Scherk--Schwarz compactification of type IIB string theory on a circle of radius $1/R$, upon appropriately modifying the T-duality rules via mass-dependent terms \cite{Bergshoeff:1996ui}. 
Given these facts, starting from the connection between massive type IIA and type IIB, and knowing how to obtain type IIB from M-theory, \cite{Hull:1998vy} obtains massive type IIA from M-theory.

The key ingredient for the Scherk--Schwarz reduction of type IIB string theory is the SL$(2,\mathbb{Z})$ duality symmetry.\footnote{The bosonic sector of type IIB supergravity has a continuous SL$(2,\mathbb{R})$ symmetry, which becomes SL$(2,\mathbb{Z})$ after imposing charge quantization. Taking into account the action on fermions, one needs to consider the metaplectic double cover, Mp$(2,\mathbb{Z})$~\cite{Dabholkar:1997zd,Pantev:2016nze}. Then, the inclusion of orientation-reversing transformations further enlarges the bosonic duality group to GL$(2,\mathbb{Z})$. Taking into account the action on bosons and fermions, the complete group is a pin$^+$ version of the double cover of GL$(2,\mathbb{Z})$ \cite{Tachikawa:2018njr}. None of these refinements affects the construction of \cite{Hull:1998vy}, since the parabolic SL$(2,\mathbb{Z})$ monodromy used there can be lifted to the full duality group.} 
Elements of the parent SL$(2,\mathbb{R})$ symmetry in supergravity can be divided into elliptic, hyperbolic, and parabolic types. The last type, which is the one relevant to constructing massive type IIA, is represented by
\begin{equation}
\label{eq:SL2_parabolic}
    \mathbb{T}[m] = \begin{pmatrix} 1&m \\ 0& 1 \end{pmatrix}\,.
\end{equation}
Under an ${\rm SL}(2,\mathbb{Z})$ transformation $g$, the complex scalar of type IIB, $\tau = C_0 + i e^{-\phi}$, transforms as
\begin{equation}
    \tau \to g \circ \tau = \frac{a\tau +b}{c \tau +d}\,, \qquad g = \left(\begin{array}{cc} a & b \\ c & d \end{array}\right)\,.
\end{equation}
Consider now type IIB string theory on an $S^1_y$ with coordinate $y\sim y+1$. We can perform a Scherk--Schwarz reduction with the ansatz
\begin{equation}
    \tau(x,y)=g(y)\circ\tau(x)\,,
\end{equation}
where $\tau(x)$ is a scalar field in the reduced theory. In general, the twist matrix $g(y)$ is not periodic but has a monodromy $g(y+1)=\mathbb{M} g(y)$. For $g(y)=\exp(\mathfrak{M}y)$, the monodromy $\mathbb{M}=g(y+1)(g(y))^{-1}$ is independent of $y$.
In the type IIB theory, $\mathbb{M} \in {\rm SL}(2,\mathbb{Z})$ and, at fixed $x$, $\tau(y)=g(y)\circ\tau$ locally specifies the complex structure of an auxiliary torus fibered over the circle.
Choosing periodic local coordinates $z_{1,2} \sim z_{1,2} +1$ on this auxiliary torus, the resulting three-dimensional fibration has metric
\begin{equation}
\label{eq:T2bundleoverS1}
    \dd s^2 = R^2 \dd y^2 + \frac{A}{{\rm Im}\tau(y)}|\dd z_1 + \tau(y) \dd z_2|^2 \,,
\end{equation}
where $A$ is the area of the torus. Note that $\tau$ may depend on other coordinates, but we are simply indicating its $y$-dependence because this is the relevant one for the Romans mass.
The special case that yields massive type IIA is
\begin{equation}\label{eq:parabolic_monodromy}
    g(y) = \left(\begin{array}{cc}1 & m y\\ 0 & 1 \end{array}\right)\,, \qquad \mathfrak{M} = \left(\begin{array}{cc} 0 & m \\ 0 & 0 \end{array}\right)\,,   \qquad \mathbb{M} \equiv \mathbb{T}[m] = \left(\begin{array}{cc} 1 & m \\ 0 & 1 \end{array}\right)\,,
\end{equation}
such that 
\begin{equation}
\label{eq:mIIA_tau}
    \tau (x,y) = \tau(x) + m\, y\, .
\end{equation}
Although $\tau(y)$ is not periodic as a complex-valued function, the metric is globally defined on $S_y^1$. Indeed, going once around $S^1_y$, which results in $\tau(y) \to \tau(y) + m$, is accompanied by a large diffeomorphism of the torus fiber,
\begin{equation}
    z_1\to z_1-m z_2 \,,
\label{eq:parabolic_large_diffeomorphism}
\end{equation}
so that $\dd z_1 + \tau (y) \dd z_2$ is invariant. 
Therefore, \eqref{eq:parabolic_large_diffeomorphism} is the transition function that glues the torus fibers around $S_y^1$. Since $m$ is an integer, it preserves the torus lattice and defines an integral parabolic monodromy.
From the real part of the ansatz \eqref{eq:mIIA_tau}, one can see that $C_0(x,y) = C_0(x) + m\, y$, and therefore the monodromy shifts the axion $C_0$ by $m$.\footnote{Here and in section \ref{sec:mpmfromM}, we work in the background truncation in which the metric Kaluza--Klein vector associated with the $y$-circle vanishes, $A_{\rm KK}^{(y)}=0$. More generally, defining the invariant one-form $\eta_y=\dd y+A_{\rm KK}^{(y)}$, one has $ \dd\widehat C_0 = \dd C_0-mA_{\rm KK}^{(y)}+m\eta_y$. The connection $A_{\rm KK}=my\,\dd z_2$ introduced below is, instead, the connection of the $S^1_{z_1}$ bundle and should not be confused with $A_{\rm KK}^{(y)}$.} 
Since $\mathbb{M} \in {\rm SL}(2,\mathbb{Z})$, $m \in \mathbb{Z}$ is quantized, which will correspond to the quantization of the Romans mass. 
To see an explicit example of \eqref{eq:T2bundleoverS1}, let $\tau_0$ be the background value of $\tau(x)$ and take $\tau_0=i$, $A=R=1$. The metric assumes the simple form
\begin{equation}
\label{eq:3DNilMan}
    \dd s^2 = \dd y^2 + (\dd z_1 + m\,y\, \dd z_2)^2 + \dd z_2^2\,,
\end{equation}
which is that of a three-dimensional Nilmanifold. 

To finally obtain massive type IIA, let $B(A,R)$ be the torus bundle over the circle of radius $R$ and coordinate $y$, with the torus modulus given by $\tau(y) = \tau_0 + my$. In the limit of vanishing area, $A \to 0$, M-theory on $B(A,R)$ is dual to type IIB string theory on a circle of radius $R$ with Scherk--Schwarz ansatz  \eqref{eq:mIIA_tau}. As shown in \cite{Bergshoeff:1996ui}, this type IIB setting is T-dual to massive type IIA compactified on a circle of radius $1/R$. 
Hence, to obtain the ten-dimensional massive type IIA theory, we need to take the limit $R\to 0$. Putting everything together, M-theory on $B(A,R)$ in the limit $A\to 0$, $R \to 0$ gives rise to massive type IIA string theory \cite{Hull:1998vy}.

Before proceeding, we want to stress that the construction of~\cite{Hull:1998vy} is to be understood as a local picture: if one wants to obtain a solution to the equations of motion, $\tau$ and $R$ in \eqref{eq:T2bundleoverS1} must have some additional spacetime dependence. 
In fact, in this way one can connect M-theory on $B(A,R)$ with the Polchinski--Witten solution~\cite{Polchinski:1995df}, reading the former as a description at a fixed value of the coordinate transverse to the D8-branes. The precise correspondence can be found in~\cite{Hull:1998vy} after eq.~(23) and consists of taking a D8-brane in type IIA and T-dualizing it along a worldvolume direction to get a smeared D7-brane in type IIB. Then, the axio-dilaton becomes as in~\eqref{eq:mIIA_tau}, where the coordinate $x$ is orthogonal to the original D8 and its dependence is encoded in a one-dimensional harmonic function. This is consistent with recent results in the framework of dynamical cobordism~\cite{Buratti:2021fiv, Angius:2022aeq}, where a critical exponent indicates that massive type IIA is inconsistent in the absence of O8/D8-branes~\cite{Calderon-Infante:2026ymy,Makridou:2026jzy}, and thus in the absence of an embedding in a global solution with sources.
In this work, we will be mainly interested in the local description, which corresponds to taking $x$ fixed in the above conventions, focusing on the mechanism that produces the Romans mass from M-theory, rather than on a global solution with non-vanishing Romans mass. 

There is an interesting feature of the space $B(A,R)$ that plays a role in this construction. $B(A,R)$ is defined as a torus bundle over a circle. However, such a bundle is also a circle bundle over a torus when the monodromy $\mathbb{M}\in {\rm SL}(2,\Z)$ is unipotent~\cite[Exercise~11.4.11]{Martelli:geomtop}, which includes the parabolic case that we are considering. This explains why we found a Nilmanifold in \eqref{eq:3DNilMan}, because every orientable circle bundle over a torus is a Nilmanifold. From this perspective, the Euler number of the $S^1$ bundle over $T^2$ becomes $m$ in the monodromy matrix in \eqref{eq:parabolic_monodromy}.
Let us verify this explicitly in the case of \eqref{eq:3DNilMan}. This space can be viewed as a circle bundle over $T^2$, with fiber coordinate $z_1$ and base coordinates $(y,z_2)$. The one-form
\begin{equation}
    \eta^1=\dd z_1+A_{\rm KK}
\end{equation}
is globally defined on the total space, although $A_{\rm KK}=my\,\dd z_2$ is only a locally defined Kaluza--Klein connection on the base. Still, its curvature is globally defined and given by
\begin{equation}
    F_{\rm KK}=\dd A_{\rm KK}=m\,\dd y\wedge \dd z_2\,.
\end{equation}
This circle bundle is characterized by its first Chern class, equivalently its Euler class, represented by $c_1=[F_{\rm KK}]\in H^2(T^2, \mathbb{Z})$, such that\footnote{Recall that our coordinates have unit period. With coordinates of period \(2\pi\), the same statement would read $c_1=[F_{\rm KK}/(2\pi)] \in H^2(T^2, \mathbb{Z})$.}
\begin{equation}
    \int_{T^2}c_1=m\,.
    \label{eq:c1m}
\end{equation}
Hence, the parabolic monodromy $m \in \mathbb{Z}$ is the first Chern number of the circle bundle, or its geometric flux~\cite{Lavrinenko:1997qa}. 

The fact that the space $B(A,R)$ can be viewed as an $S^1$ bundle over $T^2$ implies that the construction of \cite{Hull:1998vy} can be viewed as an ordinary M-theory reduction on the fiber $S^1$, followed by two T-dualities along the base torus. In fact, letting $z_1$ be the fiber coordinate and $y$ and $z_2$ the torus coordinates, upon reducing along $z_1$, the Kaluza--Klein connection becomes the type IIA RR one-form $C_1$ while the geometric flux becomes its two-form field strength, $F_2=m\,\dd y\wedge \dd z_2$. T-duality along $z_2$ leads to type IIB with RR one-form flux $F_1=m\,\dd y$. Since $F_1=\dd C_0$, this gives $C_0(x,y)=C_0(x)+my$ locally, precisely as in~\eqref{eq:mIIA_tau}. To reach massive type IIA, we perform a second T-duality along $y$, under which $F_1$ becomes the zero-form flux $F_0=m$. 
To summarize, after reducing M-theory along $z_1$, one performs two T-dualities: the first, along $z_2$, is part of the standard duality between M-theory on $T^2$ and type IIB string theory; the second, along $y$, maps the Scherk--Schwarz compactification of type IIB to massive type IIA. 

Identifying the Romans mass with the Chern number \eqref{eq:c1m}, after T-duality, also allows us to understand a further refinement due to \cite{Moore:2002cp}. The K-theory formulation of RR fields implies the existence of a topological phase in the supergravity effective action~\cite{Diaconescu:2000wy} that is not invariant under T-duality. In the notation of \cite{Moore:2002cp}, the construction of \cite{Hull:1998vy} trades the geometric flux $n_1\equiv \int_{T^2}F_2$ for the Romans mass $n_0\equiv F_0$, and the RR phase is not symmetric under the exchange $n_0\leftrightarrow n_1$. Therefore, the proposal of \cite{Moore:2002cp} is that the construction should be viewed as a \emph{quantum equivalence}, relating sums over type IIA and M-theory field configurations rather than individual configurations. A similar qualification already arises for M-theory on $M_{10}\times S^1$, where the correspondence relates sums over K-theory lifts on the type IIA side to sums over torsion shifts of the four-form flux on the M-theory side \cite{Diaconescu:2000wy}. With a non-zero Romans mass, the required sum also involves non-torsion four-form flux sectors, described in~\cite{Moore:2002cp} by a sum over $E_8$ bundles on $M_8\times B$.

\section{Massless type 0A from M-theory}\label{sec:0A_M-theory}

In this section, we review some properties of type 0 string theories and the recent conjecture of \cite{Baykara:2026gem}, connecting massless type 0A string theory and M-theory. This will fix our conventions for later sections. 

\subsection{Type 0 strings}
\label{sec:type0st}

Ten-dimensional type 0 theories are closed and oriented string theories that are non-supersymmetric counterparts of the type II theories. There are two types of type 0 string theories, depending on the chiralities of the left- and right-moving Ramond ground states: opposite (A) and equal (B). Both have purely bosonic spectra. 

From the worldsheet perspective, type 0 theories are obtained by imposing a diagonal GSO projection: one assigns the same spin structures to the left- and right-moving fermions instead of summing independently over the two spin structures as in type II. Their torus partition functions can be written as \cite{Kaidi:2019tyf}
\begin{equation}
    \mathcal{Z}^{(n)} = \frac{1}{2} \sum_{\sigma} (-1)^{n\,{\rm Arf}(T^2, \sigma)} \mathcal{Z}(\sigma)\,  \overline{\mathcal{Z}(\sigma)}\,, \qquad
    n= \begin{cases} 1 & \text{type 0A}\,,\\ 0 & \text{type 0B}\,, \end{cases}
\end{equation}
where the sum runs over the spin structures $\sigma$ on the torus, ${\rm Arf}(T^2,\sigma)$ is the corresponding Arf invariant, and $\mathcal{Z}(\sigma)$ and $\overline{\mathcal{Z}(\sigma)}$ are the left- and right-moving worldsheet partition functions.
The diagonal GSO projections and modular invariance select the sectors
\begin{eq}
    \text{type 0A}: \qquad& ({\rm NS}^{+},{\rm NS}^{+})\,,\quad ({\rm NS}^{-},{\rm NS}^{-})\,,\quad ({\rm R}^{+},{\rm R}^{-})\,,\quad ({\rm R}^{-},{\rm R}^{+})\,,\\
    \text{type 0B}: \qquad & ({\rm NS}^{+},{\rm NS}^{+})\,,\quad ({\rm NS}^{-},{\rm NS}^{-})\,,\quad ({\rm R}^{+},{\rm R}^{+})\,,\quad  ({\rm R}^{-},{\rm R}^{-})\,,
\end{eq}
where ${\rm NS}^{\pm}$ and ${\rm R}^{\pm}$ denote positive and negative $\Z_2$-parity with respect to the left- and right-moving worldsheet fermion number $(-1)^{G_{L,R}}$.
The mixed NS-R and R-NS sectors are absent, so that the perturbative spectrum does not contain spacetime fermions. The $({\rm NS}^{-},{\rm NS}^{-})$ ground state gives rise to a tachyon $\mathcal{T}$, while the massless NS-NS spectrum coincides with that of the type II theories. Instead, the RR spectrum is doubled; see Table~\ref{tab:type0-closed-spectrum}.

\begin{table}[ht]
\centering
\renewcommand{\arraystretch}{1.15}
\begin{tabular}{c|c|c}
Sector & Type 0A & Type 0B\\
\hline
$({\rm NS}^{+},{\rm NS}^{+})$
    & $g_{\mu\nu},\,B_{\mu\nu},\,\phi$
        & $g_{\mu\nu},\,B_{\mu\nu},\,\phi$ \\
$({\rm NS}^{-},{\rm NS}^{-})$
    & $\mathcal{T}$
        & $\mathcal{T}$ \\
$({\rm R}^{+},{\rm R}^{-})$
    & $C_1,\,C_3$
        & --- \\
$({\rm R}^{-},{\rm R}^{+})$
    & $C'_1,\,C'_3$
        & --- \\
$({\rm R}^{+},{\rm R}^{+})$
    & ---
        & $C_0,\,C_2,\,C_4$ \\
$({\rm R}^{-},{\rm R}^{-})$
    & ---
        & $C'_0,\,C'_2,\,C'_4$ \\
\end{tabular}
\caption{Perturbative closed string spectrum of the type 0 theories. The two RR four-forms of type 0B give rise to a self-dual and an anti-self-dual five-form field strength. Together, they can be packaged into a single five-form field strength with no self-duality constraint.}
\label{tab:type0-closed-spectrum}
\end{table}

The doubling of the RR fields implies a corresponding doubling of the charged D-branes; see~\cite{Klebanov:1998yya,Bergman:1999km,Dudas:2001wd}.
Introducing the canonically normalized potentials
\begin{align}
    C_{p+1}^{\pm} = \frac{1}{\sqrt{2}}\left(C_{p+1}\pm C'_{p+1}\right)\,,
\end{align}
for every allowed value of $p$, there are two types of branes, denoted by ${\rm D}p^+$ and ${\rm D}p^-$, which couple to $C_{p+1}^+$ and $C_{p+1}^-$ respectively. Each type has a corresponding anti-brane carrying the opposite RR charge. We summarize the D-brane spectrum in Table~\ref{tab:type0-Dbranes}.
\begin{table}[ht]
\centering
\renewcommand{\arraystretch}{1.15}
\begin{tabular}{c|c|c}
& charged  & uncharged (unstable) \\
\hline
Type 0A & ${\rm D}p^{\pm}$,\,\, $p$ even & $\widehat{{\rm D}p}^{\pm}$,\,\, $p$ odd \\
Type 0B & ${\rm D}p^{\pm}$,\,\, $p$ odd & $\widehat{{\rm D}p}^{\pm}$,\,\, $p$ even \\
\end{tabular}
\caption{D-branes of type 0 theories. Charged branes couple to the doubled potentials $C_{p+1}^\pm$ and are free of open-string tachyons. Hatted branes are not charged under RR fields and have open-string tachyons; see~\cite{Dudas:2001wd,Kaidi:2019tyf}.}
\label{tab:type0-Dbranes}
\end{table}
Open strings stretching between branes of the same type have a purely bosonic spectrum, whereas strings stretching between a ${\rm D}p^+$ and a ${\rm D}p^-$ are purely fermionic.
The two types of branes have opposite linear couplings to the closed-string tachyon~\cite{Klebanov:1998yya} and, at $\mathcal{T}=0$, the charged ones have tension $T_{{\rm D}p^\pm}^{(0)}=T_{{\rm D}p}^{({\rm II})}/\sqrt 2$.
The Wess--Zumino coupling is
\begin{equation}
    S_{\rm WZ} = \mu_p\int\left(q\,C_{p+1}+ q'\,C'_{p+1}\right) = \sqrt{2}\,\mu_p\int \left(q_+C_{p+1}^+ +q_-C_{p+1}^-\right)\,,
\end{equation}
where $ q_\pm=\frac{q\pm q'}{2}$ and $\mu_p=\mu_p^{({\rm II})}/\sqrt{2}$. Here, $T_{{\rm D}p}^{({\rm II})}$ and $\mu_p^{({\rm II})}$ refer to the type II quantities, with $\mu_p^{({\rm II})}=g_s T_{{\rm D}p}^{({\rm II})}=(2\pi)^{-p}(\alpha')^{-\frac{p+1}{2}}$. Hence, in these conventions, the ${\rm D}(-1)^\pm$-branes are minimally charged under the axions ${C_0^\pm}$, with $C_0^\pm \sim C_0^\pm +1$. 
As in type II theories, T-duality exchanges type 0A and type 0B, acting separately on the two diagonal RR sectors \cite{Meessen:2001wk}. 

To write the low-energy actions of the type 0 theories, it is convenient to introduce the RR polyforms
\begin{align}
    C^\pm_{A}&=C_1^\pm+C_3^\pm+C_5^\pm+C_7^\pm+C_9^\pm \,,\\ 
    C^\pm_{B}&=C_0^\pm+C_2^\pm+C_4^\pm+C_6^\pm+C_8^\pm \,.
\end{align}
For example, in type 0A, the two charged D8-branes couple electrically to the nine-form potentials $C_9^\pm$, whose field strengths are dual to the two RR zero-form field strengths, up to tachyon-dependent factors~\cite{Klebanov:1998yya,Bergman:1999km,Meessen:2001wk}.
Away from D8 sources, the Bianchi identities $\dd F_0^\pm=0$ imply that the zero-form field strengths are constant, and we parametrize them as
\begin{equation}
    F_0^+ = m_+\,, \qquad F_0^- = m_-\,.
\end{equation}
In general, the gauge-invariant RR field strengths are
\begin{equation}
    F^\pm=\dd C^\pm-H_3\wedge C^\pm+ m_\pm e^{B_2}\,,
\end{equation}
where $C^\pm$ is either $C^\pm_A$ or $C^\pm_B$, and in the latter case, the last term is not present. In type 0A, $m_+$ and $m_-$ deform the two RR sectors in the same way that the Romans mass enters the type IIA RR sector; therefore, they can be regarded as the two type 0A analogs of the type IIA Romans mass.
The type 0 action in the democratic formulation is~(in the mostly-minus signature of \cite{Meessen:2001wk})
\begin{equation}
\label{eq:Sdemtype0}
    S_{\rm{dem}} =\int \dd^{10}x\,\sqrt{|g|}\bigg\{ e^{-2\phi}\bigg[ R-4(\partial\phi)^2 +\frac12 |H_3|^2 +\frac12(\partial\mathcal{T})^2 -V(\mathcal{T}) \bigg] +\mathcal{L}_{RR} \bigg\}\, ,
\end{equation}
with
\begin{equation}
    \mathcal{L}_{RR} = \left\{ \begin{array}{cc} -\frac12 f_+(\mathcal{T}) \sum_{n=0}^{5}|F_{2n}^+|^2 & \qquad \text{type 0A}\\ \frac12 f_+(\mathcal{T}) \sum_{n=0}^{4}|F_{2n+1}^+|^2 & \qquad \text{type 0B} \end{array}\right.
\end{equation}
and $|F_p|^2 = \frac{1}{p!}F_{\mu_1\dots \mu_p}F^{\mu_1 \dots \mu_p}$. The tachyon coupling must satisfy \cite{Klebanov:1998yya, Meessen:2001wk}
\begin{equation}
    f_\pm(\mathcal{T})=e^{\pm h(\mathcal{T})}\,, \qquad h(-\mathcal{T})=-h(\mathcal{T})\,,
\end{equation}
where $h(\mathcal{T})$ is an odd function of $\mathcal{T}$. The appearance of only RR fields of the $+$ type in \eqref{eq:Sdemtype0} is due to the democratic formulation of type 0 theories (whether $+$ or $-$ fields appear is a convention), which follows from the relation $F_{10-2n}^\pm=(-1)^{n+1}f_\mp(\mathcal{T})* F_{2n}^\mp$ for type 0A and $F_{9-2n}^\pm=(-1)^n f_\mp(\mathcal{T})* F_{2n+1}^\mp$ for type 0B; therefore, no duality constraint has to be imposed on the $F^+$ field strengths in \eqref{eq:Sdemtype0}.

Note that $m_+$ and $m_-$ are independent deformation parameters, but they might be related on a particular background. For example, for a constant tachyon profile, $\mathcal{T}=\mathcal{T}_0$, at an extremum of its potential, $V'(\mathcal{T}_0)=0$, the
tachyon equation of motion gives
\begin{equation}
    \sum_n\frac{(-1)^n}{2} \left[ f_+(\mathcal{T}_0)|F_n^+|^2 -f_-(\mathcal{T}_0)|F_n^-|^2 \right]=0\,,
\end{equation}
provided $h'(\mathcal{T}_0)\neq0$. If $F_0^\pm= m_\pm$ are the only non-vanishing RR fluxes and the tachyon vanishes, $\mathcal{T}_0=0$, this condition reduces to
\begin{equation}
    m_+^2=m_-^2\,.
    \label{eq:T0=0}
\end{equation}
Consequently, unequal values of $|m_+|$ and $|m_-|$ source the closed-string tachyon and require either a non-vanishing tachyon profile or additional RR background fluxes. 

The type 0 theories can also be obtained as ten-dimensional orbifolds of the corresponding type II theories, taking the quotient by the spacetime fermion parity:
\begin{align}
    \text{type 0A} = \frac{\text{type IIA}}{(-1)^{F}}\,, \qquad \text{type 0B} = \frac{\text{type IIB}}{(-1)^{F}}\,,
\end{align}
where $F$ denotes the spacetime fermion number. This gives another way to see that type 0 theories are purely bosonic in spacetime. 
A basic property of orbifolds is that the orbifolded theory has an additional symmetry, which is known as the quantum symmetry of the orbifold. If the orbifold group $\Gamma$ is abelian, its quantum symmetry group is isomorphic to $\Gamma$; see, for example, \cite{Ginsparg:1988ui,Robbins:2021ibx}. The type 0 theories have a $\mathbb{Z}_2$ quantum symmetry,
\begin{equation}
\label{eq:Qsymm}
    {Q}=(-1)^{G_L}\,,
\end{equation}
where $G_L$ is the left-moving worldsheet fermion number. Orbifolding them by this quantum symmetry recovers the parent type II theories.
$Q$ acts trivially on the untwisted fields and with a minus sign on the twisted fields; in particular,
\begin{equation}
    {Q}:\qquad \mathcal{T}\to -\mathcal{T}\,, \qquad C_p\to C_p\,, \qquad C'_p\to -C'_p\,, \qquad C_p^\pm \to C_p^\mp\,.
\end{equation}
As we now review, the proposal of \cite{Baykara:2026gem} can be understood as a ``geometrization'' of the quantum symmetry of type 0A string theory.

\subsection{Massless type 0A as M-theory on \texorpdfstring{$S^1 \vee S^1$}{S1vS1}}
\label{sec:0AfromMtheory}

Type II string theories can be obtained from M-theory through circle or torus compactifications. Type 0 theories cannot be obtained analogously, at least not without additional ingredients. For example, it is unclear how to generate the doubling of RR fields geometrically; see~\cite{Bergman:1999km,Costa:2000nw,Russo:2001tf} for some initial attempts.
Motivated by this difficulty and inspired by the doubled RR sector, it was recently conjectured in~\cite{Baykara:2026gem} that type 0 theories arise from ``quantum compactifications'' of M-theory on singular spaces. In particular, type 0A would arise from a one-dimensional space with topology $S^1_+\vee S^1_-$, the wedge sum of two circles labeled $+$ and $-$, which we call wedge factors. These compactifications are unconventional: the internal space carries additional structure, shifting the focus from the space itself to the functions defined on it, as we now review.

The space $S^1_+\vee S^1_-$ is singular at the junction point $p$ where the two circles meet. Equivalently, it can be represented as a circle with two distinct points $p_+$ and $p_-$ identified; that is, there is a homeomorphism
\begin{equation}\label{eq:p_pm_def}
    S^1_+\vee S^1_- \cong S^1/\{p_+\sim p_-\}\, .
\end{equation}
The singularity at the junction can be resolved in two topologically inequivalent ways:
\begin{align}
    \label{eq:CRP}
    S_+^1 \vee S_-^1 &\,\,\to\,\, S^1\, ,\\
    \label{eq:DRP}
    S_+^1 \vee S_-^1 &\,\,\to\,\, S_+^1 \sqcup S_-^1\,.
\end{align}
The first resolution is connected, whereas the second is disconnected. The connected resolution admits two choices depending on the relative orientation of the two circles.

The proposal of~\cite{Baykara:2026gem} is that both resolutions are simultaneously realized in M-theory, with different components of the eleven-dimensional fields experiencing different resolutions. Consequently, no single geometry or topology is common to all fields. The topological space $S^1_+\vee S^1_-$ alone is therefore insufficient to define this ``quantum compactification'': one must also specify how each field component behaves at the junction.
To describe the possible matching conditions, consider a function $f$ on the circle before identifying $p_+$ and $p_-$, defined as in~\eqref{eq:p_pm_def}. Assuming that $f$ is continuous near these points, continuity at the junction requires $f(p_+)=f(p_-)$. If $f$ is smooth near both points, we additionally impose the smoothness condition $\partial^k f(p_+)=\epsilon^k\partial^k f(p_-)$ for every $k\geq1$, where $\epsilon=+1$ or $-1$ according to whether the identification preserves or reverses the local orientation.

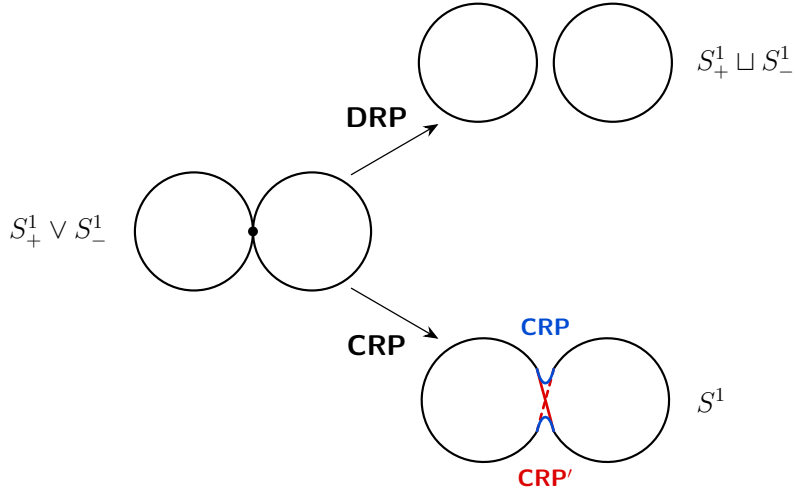
\begin{figure}[ht]
\centering
\resizebox{.7\linewidth}{!}{%
\begin{tikzpicture}[
  x=1cm,
  y=1cm,
  loop/.style={
    draw=black,
    line width=1.1pt,
    line cap=round,
    line join=round
  },
  resolution arrow/.style={
    draw=black,
    line width=0.65pt,
    -{Stealth[length=2.7mm,width=2.1mm]},
    line cap=round
  },
  transition label/.style={
    font=\sffamily\bfseries\Large
  },
  space label/.style={
    font=\Large
  },
  smoothing/.style={
    line width=1.35pt,
    line cap=round,
    line join=round
  }
]
\definecolor{crpred}{RGB}{220,0,0}
\definecolor{crpblue}{RGB}{0,75,210}
\draw[loop] (-1.05,0) circle[radius=1.05];
\draw[loop] ( 1.05,0) circle[radius=1.05];
\node[space label,anchor=east]
  at (-2.45,0)
  {$S^1_{+}\vee S^1_{-}$};
\fill[black] (0,0) circle[radius=2.5pt];
\draw[loop] (4.00,3.00) circle[radius=1.05];
\draw[loop] (6.40,3.00) circle[radius=1.05];
\node[space label,anchor=west]
  at (7.75,3.00)
  {$S^1_{+}\sqcup S^1_{-}$};
\begin{scope}[shift={(5.20,-3.00)}]
  \draw[loop]
    (-0.147,0.550)
    arc[start angle=30,end angle=330,radius=1.10];
  \draw[loop]
    (0.147,0.550)
    arc[start angle=150,end angle=-150,radius=1.10];
  \draw[smoothing,draw=crpred]
    (-0.147,0.550)
    .. controls (-0.060,0.20) and (0.060,-0.20) ..
    (0.147,-0.550);
  \draw[smoothing,draw=crpred,dashed]
    (-0.147,-0.550)
    .. controls (-0.060,-0.20) and (0.060,0.20) ..
    (0.147,0.550);
  \draw[smoothing,draw=crpblue]
    (-0.147,0.550)
    .. controls (-0.070,0.22) and (0.070,0.22) ..
    (0.147,0.550);
  \draw[smoothing,draw=crpblue]
    (-0.147,-0.550)
    .. controls (-0.070,-0.22) and (0.070,-0.22) ..
    (0.147,-0.550);
  \node[
    font=\sffamily\bfseries\large,
    text=crpblue
  ] at (0,1.30) {CRP};
  \node[
    font=\sffamily\bfseries\large,
    text=crpred
  ] at (0,-1.38) {CRP$'$};
\end{scope}
\node[space label,anchor=west]
  at (7.75,-3.00)
  {$S^1$};
\draw[resolution arrow]
  (1.75,1.01) -- (3.30,1.91);
\draw[resolution arrow]
  (1.75,-1.01) -- (3.30,-1.91);
\node[transition label] at (2.20,2.03) {DRP};
\node[transition label] at (2.20,-2.03) {CRP};
\end{tikzpicture}%
}
\caption{DRP and CRP types of resolutions of $S^1_{+}\vee S^1_{-}$.}
\label{fig:drp-ssp-resolutions}
\end{figure}

Fields that extend smoothly to the connected resolution \eqref{eq:CRP} are said to have the Connected Resolution Property (CRP), while fields that extend to the disconnected resolution \eqref{eq:DRP} are said to have the Disconnected Resolution Property (DRP). DRP fields are independently periodic on each circle, with no matching condition at the junction, while CRP fields extend smoothly across the junction in the connected resolution. These conditions are depicted in Figure~\ref{fig:drp-ssp-resolutions}. Fields compatible with both resolutions are said to have the Strong Smoothness Property (SSP). The connected resolution can be realized in two ways, depending on whether the relative orientation of the two circles is reversed across the junction or not. When it is reversed, we denote the corresponding property by CRP$'$; similarly, one obtains the orientation-reversed SSP$'$.

The derivatives in the smoothness condition must be taken with respect to the angular coordinates $\theta_\pm$ on the two circles. In fact, if derivatives were defined in terms of the arc-length, the length scales $R_\pm$ associated with $S^1_\pm$ would enter the smoothness condition explicitly. Indeed, expanding the functions on the individual wedge factors as $f_\pm=\sum_{n_\pm\in\mathbb{Z}}f_{n_\pm}e^{in_\pm\theta_\pm}$, matching individual Fourier modes using arc-length derivatives would require $n_-=\pm ( R_-/R_+) \, n_+$. For generic radii, this is incompatible with the quantization of $n_\pm$. Using angular derivatives, smoothness for each individual Fourier mode instead gives $n_-=\pm n_+$, independently of $R_\pm$.
Accordingly, in the prescription of~\cite{Baykara:2026gem}, the SSP and SSP$'$ conditions correlate the Kaluza--Klein excitations on the two circles, yielding quantum numbers $(n,n)$ and $(n,-n)$, respectively. DRP fields instead admit independent Kaluza--Klein modes, labeled by $(n,0)$ and $(0,n)$.\footnote{Strictly speaking, since the fields do not all experience the same internal geometry, there is no conventional Kaluza--Klein reduction. When discussing Kaluza--Klein excitations, one is really considering the Fourier expansion of the fields. The focus is again on the space of functions rather than on the geometry.}

In the quantum compactification, the components of the bosonic M-theory fields are taken to be either DRP or SSP/SSP$'$. The eleven-dimensional Majorana gravitino $\Psi_\mu$, which is the only fermionic field, is instead assigned a modified SSP condition denoted by SSP$^*$. This is an SSP condition with an additional minus sign in the identification at the junction. In the prescription of \cite{Baykara:2026gem}, when $p_+$ and $p_-$ are identified with the junction $p$, the condition $\Psi_\mu(p_+)=-\Psi_\mu(p_-)$ is implemented as $\Psi_\mu(p)=0$.\footnote{This should not be confused with an ordinary antiperiodic spin structure on a circle, $\Psi_\mu(\theta+2\pi)=-\Psi_\mu(\theta)$, which gives rise to half-integer Kaluza--Klein momenta $n+\frac12$. The SSP$^*$ condition is instead an odd identification at the junction: the modes remain periodic on each circle and therefore carry integer Kaluza--Klein quantum numbers.} This removes the constant mode, which would give rise to two ten-dimensional Majorana--Weyl gravitini of opposite chirality together with the associated spin-$\frac12$ dilatini in an ordinary circle reduction.
The absence of this zero mode is interpreted as the absence of fermions in the perturbative ten-dimensional spectrum, as required to match type 0A. In addition, the two connected resolutions CRP and CRP$'$ are assumed to be correlated with the two eigenvalues of the ten-dimensional chirality operator $\Gamma_{11}$. The $\Gamma_{11}=+1$ component is associated with CRP and has Kaluza--Klein quantum numbers $(n,n)$, whereas the $\Gamma_{11}=-1$ component is associated with CRP$'$ and has Kaluza--Klein quantum numbers $(n,-n)$; as we just explained, $n=0$ is excluded by the SSP$^*$ condition. 

To summarize, the quantum compactification is specified by the following conditions on the field space:
\begin{align}
\text{DRP}: \quad &\text{periodic on each circle, i.e., } f_{\pm}(\theta_{\pm}+2\pi)=f_{\pm}(\theta_\pm);\nonumber \\
\text{CRP}: \quad &\text{smooth across the junction, i.e.,} \ \partial_{\theta_+}^kf_+(p_+) = \partial_{\theta_-}^kf_-(p_-), k\geq 0; \nonumber \\
 \text{CRP$'$}: \quad &\text{smooth across the junction, i.e., } \partial_{\theta_+}^kf_+(p_+) = (-1)^k \partial_{\theta_-}^kf_-(p_-), k\geq 0; \nonumber \\
\text{SSP/SSP$'$}: \quad &\text{periodic on each circle and smooth across the junction}; \nonumber \\
\text{SSP$^*$}: \quad &\text{periodic on each circle, smooth and odd under the junction identification}. \nonumber
\end{align}

There is a further condition to be imposed on the space of functions that has to do with the junction.
As we mentioned above, one of the motivations behind the choice of the topological space underlying the proposal of~\cite{Baykara:2026gem} is the doubling of the RR fields. In particular, type 0A string theory has a $\mathrm{U}(1)_+\times\mathrm{U}(1)_-$ gauge symmetry associated with the two RR fields $C_1^\pm$. One would like to obtain this gauge symmetry ``geometrically'' in the M-theory picture, from the isometries of the two circles $S^1_\pm$. However, the junction point in the wedge sum breaks the (continuous part of the) independent $\mathrm{U}(1)_+\times\mathrm{U}(1)_-$ rotations, so one cannot argue for two independent massless gauge fields in the resulting ten-dimensional theory. To address this, recall that the focus is on the space of functions: imposing additional conditions on this space can recover the two $\mathrm{U}(1)$ symmetries, at the cost of introducing further rules for the compactification on the singular space.
To this end, let $\alpha$, $\beta$ be arbitrary independent shifts along the two circles normalized to unit period, $\alpha\sim\alpha+1$ and $\beta\sim\beta+1$. These are associated with the two abelian gauge parameters. On the CRP function space, one imposes the following equivalence relation:
\begin{align}
    f_+(\theta_+)\sim f_+(\theta_++2\pi\alpha)\,, \qquad f_-(\theta_-)\sim f_-(\theta_-\pm2\pi\beta)\,,
\label{eq:BDVequivalence}
\end{align}
where the two signs correspond to the two possible relative orientations, CRP and CRP$'$. This additional rule restores the full $\mathrm{U}(1)_+\times\mathrm{U}(1)_-$ symmetry at the level of the space of allowed functions: intuitively, the junction point corresponds to a quantum superposition of all the points on the two circles.

With this understanding of the singular space $S^1_+\vee S^1_-$, the M-theory compactification of \cite{Baykara:2026gem} assigns to each component of the eleven-dimensional fields a resolution property such that one recovers the low-lying spectrum of type 0A string theory. Intuitively, DRP fields are doubled and hence must correspond to the eleven-dimensional components giving rise to the RR spectrum of type 0A, thus trading the doubling for a feature of the quantum compactification. The remaining fields must be SSP. 
We summarize the full prescription in Table~\ref{tab:BDV-field-content}. 
\begin{table}[ht]
\centering
\renewcommand{\arraystretch}{1.2}
\begin{tabular}{c|c|c}
M-theory field & Type 0A field & Resolution property
\\
\hline
$g_{\mu\nu}$ & $g_{\mu\nu}$ & SSP \\ 
    $g_{\mu+}$, $g_{\mu-}$ & $C_{\mu}^+$, $C_\mu^-$ & DRP \\
        $g_{++}$, $g_{--}$ & $\phi$, $\mathcal{T}$ & DRP \\
            $C_{\pm\mu\nu}$ & $B_{\mu\nu}$ & SSP \\
                $C_{\mu\nu\rho}$ & $C_{\mu\nu\rho}^\pm$ & DRP \\
                    $\Psi_{\mu}$ & --- & SSP$^*$
\end{tabular}
\caption{Resolution properties of eleven-dimensional fields on $S^1_+\vee S^1_-$ that yield the low-lying spectrum of type 0A string theory. In our conventions, $g_{++} = R_+^2$ and $g_{--}=R_-^2$.}
\label{tab:BDV-field-content}
\end{table}
Whenever a field is assigned the SSP resolution, the corresponding orientation-reversed SSP$'$ sector is also allowed. Thus, the fields $g_{\mu\nu}$ and $C_{\pm\mu\nu}$ (the latter giving rise to the unique ten-dimensional two-form $B_{\mu\nu}$) admit correlated Kaluza--Klein excitations with quantum numbers $(n,n)$ and $(n,-n)$ in the SSP and SSP$'$ sectors, respectively. In fact, the distinction between SSP and SSP$'$ only affects their massive Kaluza--Klein excitations: for $n=0$, the two conditions coincide and generate the same massless fields $g_{\mu\nu}$ and $B_{\mu\nu}$. 

\section{Massive type 0A from M-theory}
\label{sec:m0AfromMtheory}

In this section, we present our proposal for the M-theory origin of massive type 0A string theory, combining the realization of massive type IIA from M-theory of \cite{Hull:1998vy}, reviewed in section~\ref{sec:Hull}, with the proposal of \cite{Baykara:2026gem} to obtain massless type 0A from M-theory on $S^1\vee S^1$, reviewed in section \ref{sec:0AfromMtheory}. In the following, we sometimes restore the string scale $L_S=1/M_S=\sqrt{\alpha'}$ when it makes the presentation clearer. Before addressing massive type 0A, we review one last auxiliary construction: type 0B string theory from M-theory.

\subsection{Type 0B string theory}

As reviewed in section~\ref{sec:Hull}, obtaining massive type IIA from M-theory requires passing through type IIB. By analogy, one may expect type 0B to play a similar role in obtaining massive type 0A from M-theory. We then recall how type 0B string theory is realized in the proposal of~\cite{Baykara:2026gem}.  

Consider M-theory on the singular two-dimensional space
\begin{equation}
\label{eq:S1vS1xS1}
    \left(S^1_+\vee S^1_-\right)\times S^1_u\,,
\end{equation}
where the circles are parameterized by the unit-period coordinates $z_\pm=\theta_\pm/2\pi\sim z_\pm+1$ and $u\sim u+1$, with $R_+$, $R_-$, and $R_u$ denoting their sizes. Each wedge factor combines with $S^1_u$ to form a torus,
\begin{equation}
    T^2_\pm=S^1_\pm\times S^1_u\,,
\end{equation}
and the two tori are glued along their common cycle $S^1_u$; we denote this space by
\begin{equation}
    T^2_+\vee_{S^1_u}T^2_-\,.
\end{equation}
Recall the two RR axions of type 0B string theory introduced in section \ref{sec:type0st}, $C_0^\pm=\frac{C_0\pm C_0'}{\sqrt{2}}$, and normalized with unit periodicity, $C_0^\pm \sim C_0^\pm+1$. 
They enter the complex structures of the two tori $T^2_\pm$,
\begin{equation}
    \tau_\pm=C_0^\pm+i\frac{R_u}{R_\pm}\,,
\label{eq:tau_pm}
\end{equation}
while the corresponding areas are $A_\pm=R_uR_\pm$.  
The type 0B quantum symmetry \eqref{eq:Qsymm} exchanges the two wedge factors and, in particular,
\begin{equation}
    Q: \qquad \tau_+ \, \longleftrightarrow \tau_-\,.
\end{equation}
Accordingly, the $Q$-even combination contains the type 0B axio-dilaton, whereas the $Q$-odd combination contains the tachyon and the second RR axion. Schematically, one has that
\begin{equation}
    C_0+i e^{-\phi} \ \propto \ \tau_{+}+\tau_{-}\,, \qquad  \mathcal{T}+iC_0' \ \propto \ \frac{\tau_{+}-\tau_{-}}{\tau_{+}+\tau_{-}} \,,
\label{eq:0B_local_dictionary}
\end{equation}
which is consistent with the fact that type IIB is the orbifold of type 0B by $Q$.

M-theory on \eqref{eq:S1vS1xS1} is dual to type 0A on $S_u^1$, and T-duality on this geometric circle leads to type 0B on the circle $S^1_B$ with radius $R_B=\alpha'/R_u$. Ten-dimensional type 0B is thus obtained in the limit $R_B \to \infty$, or equivalently $R_u \to 0$. To understand this limit in the M-theory variables, recall that, due to the SSP prescription, M2-branes corresponding to the fundamental type 0A string must wrap the two wedge factors with equal winding numbers. The tension of the string is then proportional to $M_S^2 =M_P^3 (R_+ + R_-)$, where $M_P$ is the eleven-dimensional Planck mass. A fundamental type 0A string wrapped once around $S^1_u$ is mapped by T-duality to a Kaluza--Klein mode on $S^1_B$, whose mass is
\begin{equation}
    \frac{1}{R_B}=\frac{R_u}{\alpha'}= M_P^3 R_u(R_++R_-)\,.
\end{equation}
It follows that the ten-dimensional type 0B theory is recovered in the limit
\begin{equation}
    R_u\to 0 \,, \qquad R_\pm\to 0 \,, \qquad \frac{R_u}{R_\pm}\quad\text{fixed} \,,
\label{eq:0B_zero_area}
\end{equation}
hence sending $R_B\to\infty$ while keeping the complex
structures \eqref{eq:tau_pm} finite. As in the ordinary type IIB/F-theory limit, the auxiliary tori have vanishing areas, while their complex structures survive.

\subsection{Type 0B duality as a marked elliptic curve}

To perform a Scherk--Schwarz reduction as in \cite{Hull:1998vy}, we need to know the duality group of type 0B string theory. This has been briefly discussed in~\cite{Baykara:2026gem}, and below we expand on their proposal.\footnote{Actually, for the parabolic Scherk--Schwarz reduction that we will use below, it is not necessary to know the full duality group of type 0B, in the same way as it was not necessary in \cite{Hull:1998vy}. It suffices to know that the shifts $C_0^\pm\sim C_0^\pm+1$ are symmetries of the theory.} 

The presence of two complex structures $\tau_\pm$ might suggest that the duality group is isomorphic to two copies of ${\rm SL}(2,\Z)$. However, while it is true that the DRP resolution does contain two tori, their modular transformations cannot be independent symmetries of the full theory because the SSP sector only sees a single combined torus. The duality group must then contain only transformations that act compatibly for both types of resolutions. It has been proposed in~\cite{Baykara:2026gem, Kan:2026sea} that, at the $Q$-invariant locus
\begin{equation}
    \tau_+=\tau_-\equiv\tau_0 \,,  \qquad \mathcal{T}=C_0'=0\,,
\label{eq:0B_symmetric_locus}
\end{equation}
the duality group acting on $\tau_0$ is
\begin{equation}
    \Gamma_0(2)=\left\{
    \begin{pmatrix}
    a&b \\ c&d
    \end{pmatrix} \in{\rm SL}(2,\Z) \ : \ c\equiv 0 \pmod{2} \right\}\,.
\label{eq:Gamma0_2}
\end{equation}
Without invoking $S^1_+\vee S^1_-$, one can still argue that the duality group of type 0B at the $Q$-invariant locus~\eqref{eq:0B_symmetric_locus} is a level-two congruence subgroup of ${\rm SL}(2,\mathbb Z)$. In fact, type 0B can be obtained by putting type IIA on a circle with antiperiodic fermions in the zero-radius limit. In M-theory, this is a two-torus with one periodic and one antiperiodic spin structure. Its mapping-class group is the subgroup of ${\rm SL}(2,\Z)$ that preserves this spin structure, which is again a level-two subgroup. 

While not relying on $S^1_+\vee S^1_-$, the argument can only capture the duality at the $Q$-invariant locus. To describe the full duality symmetry, and in particular, to fix the integral normalization of the parabolic monodromy that we will use below, one has to ``geometrize'' also the $Q$-odd combination $\mathcal{T}+i C_0'$. To this end, we now introduce a formulation of the type 0B duality symmetry in terms of a marked elliptic curve.

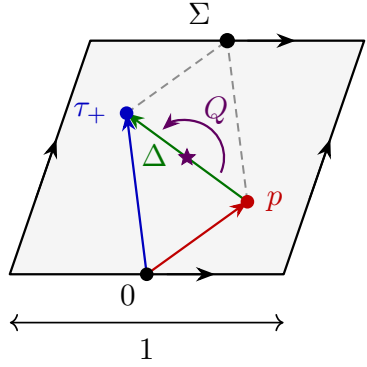
\begin{figure}[ht]
\centering
\resizebox{.33\linewidth}{!}{
\begin{tikzpicture}[
  x=1cm,
  y=1cm,
  boundary/.style={
    draw=black,
    line width=0.8pt,
    line cap=round,
    line join=round
  },
  identification/.style={
    draw=black,
    line width=.65pt,
    -{Stealth[length=2.4mm,width=1.8mm]}
  },
  tauplus/.style={
    draw=blue!75!black,
    line width=0.8pt,
    -{Stealth[length=2.2mm,width=1.7mm]}
  },
  tauminus/.style={
    draw=red!75!black,
    line width=0.8pt,
    -{Stealth[length=2.2mm,width=1.7mm]}
  },
  deltavector/.style={
    draw=green!45!black,
    line width=0.8pt,
    -{Stealth[length=2.2mm,width=1.7mm]}
  },
  auxiliary/.style={
    draw=black!45,
    densely dashed,
    line width=.7pt
  },
  point label/.style={
    font=\normalsize
  }
]
\coordinate (O)  at (0,0);
\coordinate (S)  at (1.00,2.90);
\coordinate (P)  at (1.25,0.90);
\coordinate (TP) at (-.25,2.00);
\coordinate (M)  at (0.50,1.45);

\fill[black!4]
  (-1.7,0) -- (1.7,0) -- (2.7,2.90) -- (-0.7,2.90) -- cycle;
\draw[boundary]
  (-1.7,0) -- (1.7,0) -- (2.7,2.90) -- (-0.7,2.90) -- cycle;

\draw[identification] (0.25,0) -- (0.85,0);
\draw[identification] (1.25,2.90) -- (1.85,2.90);
\draw[identification] (-1.38,0.93) -- (-1.12,1.68);
\draw[identification] ( 2.02,0.93) -- ( 2.28,1.68);

\draw[<->,line width=.6pt]
  (-1.7,-.60) --
  node[midway,below=2pt,font=\normalsize] {$1$}
  (1.7,-.60);

\draw[auxiliary] (P) -- (S);
\draw[auxiliary] (TP) -- (S);

\draw[tauminus] (O) -- (P);
\draw[tauplus] (O) -- (TP);
\draw[deltavector]
  (P) -- node[
    pos=.70,
    below left=0pt,
    inner sep=1pt,
    xshift=3pt,
    font=\normalsize,
    text=green!45!black
  ] {$\Delta$} (TP);

\draw[
  draw=violet!75!black,
  line width=.8pt,
  -{Stealth[length=2.2mm,width=1.7mm]}
]
  ([shift={(-24.25:0.46)}]M)
  arc[start angle=-24.25,end angle=131.75,radius=0.46];

\node[
  point label,
  text=violet!75!black
] at (0.86,2.03) {$Q$};

\node[
  star,
  star points=5,
  star point ratio=2.2,
  minimum size=6pt,
  inner sep=0pt,
  draw=violet!75!black,
  fill=violet!75!black
] at (M) {};

\fill[black] (O) circle[radius=2.4pt];
\node[point label,anchor=north east] at (O) {$0$};

\fill[black] (S) circle[radius=2.7pt];
\node[
  point label,
  text=black,
  anchor=south east,
  xshift=-2pt,
  yshift=2pt
] at (S) {$\Sigma$};

\fill[red!75!black] (P) circle[radius=2.4pt];
\node[
  point label,
  text=red!75!black,
  anchor=west,
  xshift=3pt
] at (P) {$p$};

\fill[blue!75!black] (TP) circle[radius=2.4pt];
\node[
  point label,
  text=blue!75!black,
  anchor=east,
  xshift=-3pt
] at (TP) {$\tau_+$};
\end{tikzpicture}
}

\caption{Marked elliptic curve $(E_\Sigma,p)$ with
$p=\tau_-$. The picture shows $\tau_+=\Sigma-p$,
$\Sigma=\tau_++\tau_-$, and $\Delta=\tau_+-\tau_-$.
The violet star represents the point $\Sigma/2$, around which
the quantum symmetry $Q$ acts as a $\pi$ rotation.}
\label{fig:marked-elliptic-curve}
\end{figure}

Away from the symmetric locus \eqref{eq:0B_symmetric_locus}, the two complex parameters $\tau_\pm$ can be naturally rearranged into the complex structure of a single elliptic curve together with a marked point. To this end, define the complex number
\begin{equation}
    p=\frac{\Sigma -\Delta}{2}=\tau_-\,, \qquad \text{with} \qquad \Sigma=\tau_+ +\tau_- \,, \qquad \Delta=\tau_+ - \tau_- \,,
\label{eq:sigma_p}
\end{equation}
and consider the elliptic curve
\begin{equation}
    E_\Sigma = \C/(\Z+\Sigma\Z)\,,
\end{equation}
of which $p\in E_\Sigma$ is a marked point and $\Sigma$ the complex structure. 
While we have chosen $p=\tau_-$, the alternative choice of $\frac{\Sigma+\Delta}{2} = \tau_+$ as the marked point is related to it by the quantum symmetry and is therefore equivalent.
Indeed, the quantum symmetry exchanges $\tau_+$ and $\tau_-$ and acts on the above variables as
\begin{equation}
    Q : \qquad \Sigma\to\Sigma \,, \qquad \Delta\to-\Delta\,, \qquad p\to \Sigma - p\sim-p\,,
\end{equation}
where the last equivalence follows because $\Sigma$ is a period of $E_\Sigma$; see Figure~\ref{fig:marked-elliptic-curve}. The local dictionary of \eqref{eq:0B_local_dictionary} becomes
\begin{equation}
    \mathcal{T}+iC_0' \ \propto \ \frac{\Delta}{\Sigma}=1-\frac{2p}{\Sigma}\,,
\end{equation}
and the $Q$-invariant locus $\mathcal{T}=C_0'=0$ corresponds to 
\begin{equation}
    p=\frac{\Sigma}{2}\,.
\end{equation}

In this formulation, we propose that the duality symmetry of type 0B string theory can be recast as the following modular action on the pair $(E_\Sigma,p)$:
\begin{equation}
    \Sigma \to \tilde\Sigma=\frac{a\Sigma+b}{c\Sigma+d}\,, \qquad p\to\tilde{p}=  \frac{p}{c\Sigma+d}\,, \qquad
    \begin{pmatrix} 
    a&b \\ c&d 
    \end{pmatrix} \in{\rm SL}(2,\Z)\,.
\label{eq:sl2z_marked_torus}
\end{equation}
Note that the action on $p$ must be supplemented by the identification
\begin{equation}
    p\sim p+r+s\Sigma\,,  \qquad r,s\in\Z \,,
\label{eq:marked_point_translations}
\end{equation}
because $p$ is a point on $E_\Sigma$.
We can now see how a level-two subgroup of ${\rm SL}(2,\Z)$ arises as the stabilizer of the marked point at the $Q$-invariant locus, $p|_{\text{$Q$-inv}}=\Sigma/2$. Indeed, acting with ${\rm SL}(2,\Z)$, the marked point becomes
\begin{equation}
    \tilde{p}=\frac{1}{2}\frac{\Sigma}{c\Sigma+d} = \frac{d\tilde\Sigma -b}{2}\,,
\label{eq:p_transformation}
\end{equation}
where the second equality follows from $ad-bc=1$. When $b$ is even, the determinant condition implies that $d$ is odd; therefore, the transformed point $\tilde p$ is equivalent to $\tilde \Sigma/2$ modulo the lattice of $E_{\tilde \Sigma}$.
To compare with~\eqref{eq:Gamma0_2}, note that at the $Q$-invariant locus, $ \Sigma=2\tau_0 = 2 p$. Hence, a transformation $\Sigma\to(a\Sigma+b)/(c\Sigma+d)$ induces $ \tau_0\to (a\tau_0+b/2)/(2c\tau_0+d)$, where $b$ is even. We thus recover $\Gamma_0(2)$ acting on $\tau_0$ as the stabilizer of the vanishing-tachyon locus.
In particular, note that the parabolic transformation $\mathbb T[2m]$ acting on $\Sigma$ induces $\mathbb T[m]$ acting on $\tau_0$. 

The key feature for us is that the existence of a single modular group does not clash with the existence of two independent axion periodicities. The two independent integral shifts
\begin{equation}
    \tau_+\to\tau_+ + m_+ \,, \qquad \tau_-\to\tau_- + m_- \,, \qquad m_\pm\in\Z 
\end{equation}
act on the variables $(\Sigma,\Delta,p)$ as
\begin{equation}
    \Sigma\to\Sigma+M \,, \qquad \Delta\to\Delta+N \,, \qquad p\to p+m_- \,,
\label{eq:marked_curve_integer_shifts}
\end{equation}
where
\begin{equation}\label{eq:M_and_N_def}
    M=m_+ + m_- \,, \qquad N=m_+ - m_- \,.
\end{equation}
These transformations are compatible with the definition of $p=(\Sigma-\Delta)/2$ since
\begin{equation}
    \frac{(\Sigma+M)-(\Delta+N)}{2} = p + m_- \sim p \,.
\end{equation}
For example, the shift $\Sigma\to\Sigma+M$ is the modular transformation $\mathbb T[M]$, while $p\to p+m_-$ is an integral translation of the marked point, which is identified with $p$ using \eqref{eq:marked_point_translations}. 

Before concluding this section, we discuss a subtle consequence of the proposed type 0B duality group: the transformation in~\eqref{eq:sl2z_marked_torus} does not always preserve the positivity of the imaginary parts of $\tau_\pm$. In fact, without loss of generality, let us parametrize
\begin{equation}
    p=\xi+\lambda \Sigma\,,\qquad \text{with}\qquad \xi,\lambda\in\R\,,
\end{equation}
so that 
\begin{equation}
    {\rm Im}\tau_-=\lambda \, {\rm Im}\Sigma\,,\qquad {\rm Im}\tau_+=(1-\lambda) \, {\rm Im}\Sigma\,.
\end{equation}
From~\eqref{eq:tau_pm}, $\lambda=\frac{R_+}{R_++R_-}$ is the relative size of one of the two wedge factors. The condition ${\rm Im}\tau_\pm>0$ is then equivalent to
\begin{equation}
    0<\lambda<1\,,
\label{eq:condition_lambda}
\end{equation}
namely, to the fact that both wedge factors have finite size. A lattice translation $p\to p+r+s\Sigma$, with $r,s\in\Z$, shifts $\lambda\to\lambda+s$. Thus, if $\lambda\notin\Z$, one can choose $s$ so that $0<\lambda+s<1$, restoring \eqref{eq:condition_lambda}. If $\lambda\in\Z$, however, every translated value remains an integer, so none lies strictly between zero and one. The boundary representatives $\lambda=0$ and $\lambda=1$ correspond to a vanishing imaginary part for one of the two $\tau_\pm$ and hence, through \eqref{eq:tau_pm}, to a limiting configuration in which one wedge factor is infinitely larger than the other. According to the conjecture of~\cite{Baykara:2026gem}, this leads to type IIB string theory. More explicitly, the transformation
in~\eqref{eq:sl2z_marked_torus} acts on $\xi$ and $\lambda$ as
\begin{equation}
    \begin{pmatrix}
        \tilde\xi \\ \tilde\lambda
    \end{pmatrix}= 
    \begin{pmatrix}
        a & -b \\ -c & d
    \end{pmatrix}
    \begin{pmatrix}
        \xi \\ \lambda
    \end{pmatrix}\,,
\end{equation}
while the identification in~\eqref{eq:marked_point_translations} becomes $(\xi,\lambda)\sim (\xi+r,\lambda+s)$. 
If $\tilde\lambda\notin\Z$, a lattice translation brings it into the interval \eqref{eq:condition_lambda}. If $\tilde\lambda\in\Z$, no such translation exists; choosing the boundary representative $\tilde\lambda=0$ or $1$ makes one of ${\rm Im}\tilde\tau_\pm$ vanish, hinting at a possible
connection with the type IIB endpoint proposed
in~\cite{Baykara:2026gem}. 
This suggests two possible interpretations:
\begin{enumerate}
    \item The transformed boundary configurations are physical and admit a type IIB interpretation, so type 0B and type IIB are connected by duality. In this case, the definition of the complex structures in \eqref{eq:tau_pm} requires a further non-geometric generalization, allowing for vanishing imaginary parts. While this is not as drastic as other non-geometric prescriptions we have already adopted, it would mean that the last remnant of a geometric formulation, the requirement of positive radii, must be relaxed.
    \item The duality group must be restricted to transformations that keep the imaginary parts of $\tau_\pm$ positive, up to lattice translations. In terms of \eqref{eq:sl2z_marked_torus}, requiring this at every point implies $c=0$. If $c\neq0$, then $|c|\geq1$, since $c$ is an integer, and the transformed $\tilde\lambda=-c\xi+d\lambda$ necessarily takes an integer value as $\xi$ varies over one period, $\xi\sim\xi+1$. Instead, if $c=0$, one has $a=d=\pm1$ and $\tilde\lambda=\pm\lambda$; therefore, starting with a non-integer $\lambda$ never leads to an integer. Hence, the only transformations that never lead to a degeneration anywhere in the positive-radius domain are $\pm\mathbb{T}[b]$. 
    At special points with $\xi=0$, the allowed set of transformations can be larger. For instance, at the $Q$-invariant locus, where $p=\Sigma/2$, one has $\lambda=1/2$ and hence $\tilde\lambda=d/2$. As found after~\eqref{eq:p_transformation}, the transformations in the stabilizer of the vanishing-tachyon locus have $b$ even. The determinant condition then implies that $d$ is odd, so $\tilde\lambda\sim1/2$ is again the $Q$-invariant value, while $\tilde\xi=-b/2\sim0$.
\end{enumerate}

Regardless of this subtlety with the interpretation, the important piece of information for our purposes is that the duality group of type 0B contains two independent axionic shift symmetries, even in the absence of two modular groups. In terms of the marked elliptic curve, one shift symmetry is encoded in the modular parameter of the elliptic curve, and the other in its marked point. 
This structure can give rise to two independent Scherk--Schwarz twists, as we describe now.

\subsection{The quantum fibration}\label{ssec:quantum_fibration}

In the construction of~\cite{Hull:1998vy} reviewed in section~\ref{sec:Hull}, massive type IIA string theory is obtained from M-theory on a $T^2$ bundle over $S^1$, denoted $B(A,R)$, in the limit $A\to 0$, $R\to0$, with a duality twist for the torus complex structure when going around the $S^1$. The structure described in the previous section suggests that massive type 0A string theory can be obtained analogously, taking the quantum compactification of M-theory on a $T^2_+\vee_{S^1_u}T^2_-$ bundle over a circle $S^1_y$, with $y \sim y +1$, in the limit of vanishing sizes and twisting along the $S^1_y$ both the complex structure $\Sigma$ and the marked point location $p$, defined as in \eqref{eq:sigma_p}. The two Romans masses of type 0A arise then from the two independent duality twists.

In fact, in the limit of zero ``area'' for $T^2_+\vee_{S^1_u}T^2_-$, this M-theory setup becomes type 0B string theory with the Scherk--Schwarz ansatz
\begin{equation}
    \Sigma(x,y)=\Sigma_0(x)+M y \,, \qquad  p(x,y) = p_0(x)+L y
\end{equation}
around $S^1_y$, where $M,L\in\Z$. From \eqref{eq:marked_curve_integer_shifts} and \eqref{eq:M_and_N_def}, this corresponds to  
\begin{equation}
    \tau_\pm(x,y) = \tau_{\pm,0}(x)+m_\pm y\,, \qquad m_\pm\in\mathbb Z \,,
\label{eq:tau_pm_winding}
\end{equation}
with $m_-=L$ and $m_+=M-L$; it therefore shifts the two RR axions $C_0^\pm$.
Note that only the real parts of $\tau_\pm$ wind, while the ratios $R_u/R_\pm$ in \eqref{eq:tau_pm} are independent of $y$ in the background. 

From the point of view of the DRP fields, \eqref{eq:tau_pm_winding} is seen as a parabolic monodromy for each wedge factor, 
\begin{equation}
    \mathcal M_\pm = \mathbb{T}[m_\pm] =
    \begin{pmatrix}
    1&m_\pm \\
    0&1
    \end{pmatrix} \in{\rm SL}(2,\Z)\,,
\label{eq:branch_monodromies}
\end{equation}
even though, as we have already emphasized, there is only a single ${\rm SL}(2,\Z)$, and the doubled monodromy comes from the marked elliptic curve structure of the conjectured type 0B duality symmetry. In fact, using \eqref{eq:sigma_p}, the ansatz \eqref{eq:tau_pm_winding} corresponds to
\begin{equation}
    \begin{aligned}
        \Sigma(x,y) & = \Sigma_0(x)+M y \,, \\
        \Delta(x,y) & = \Delta_0(x)+N y\,,\\
        p(x,y) & = p_0(x)+L y \,,
    \end{aligned} \qquad
    \begin{aligned}
        M & =m_+ + m_- \,, \\
        N & = m_+ - m_- \,, \\
        L & = m_- \,.
    \end{aligned}
\label{eq:marked_curve_winding}
\end{equation}
This means that going once around $S^1_y$, the marked curve has monodromy 
\begin{equation}
    (\Sigma,p) \to (\Sigma+M,p+L) \,.
\end{equation}
The lattices defining $E_{\Sigma+M}$ and $E_\Sigma$ coincide, as does the marked point $p+L\sim p$ (see \eqref{eq:marked_point_translations}), but the path followed around the $S^1_y$ need not be contractible. This is analogous to an axion configuration $C_0=C_{0,0}+my$ whose values at $y=0$ and $y=1$ are identified, but the configuration has a winding number $m$. 

The importance of the marked point is particularly transparent when considering the elementary windings of the two wedge factors,
\begin{align}
    (m_+,m_-)=(1,0) \quad&\to\quad (M,L)=(1,0) \,,\\
    (m_+,m_-)=(0,1) \quad&\to\quad (M,L)=(1,1) \,.
\end{align}
They both induce $\Sigma\to\Sigma+1$, and what distinguishes them is the shift of the marked point, which is trivial in the first case, while it has unit winding in the second. 

It is interesting to separate the $Q$-even and $Q$-odd components of the monodromy. The quantum symmetry exchanges the two wedge factors, acting as
\begin{equation}
    Q : \qquad (m_+,m_-)\to (m_-,m_+) \,, \qquad (M,N,L)\to (M,-N,M-L)\,.
\end{equation}
Consider first a fibration with $p_0=\Sigma_0/2$. This fibration remains on the $Q$-invariant locus when $m_+=m_-\equiv m$, in which case \eqref{eq:marked_curve_winding} gives $M=2m$, $N=0$, $L=m$; hence, $p(y)=\Sigma(y)/2$ along the entire path. In fact, in this case, one can orbifold by $Q$, obtaining type IIB string theory, where the surviving type IIB modulus transforms as
\begin{equation}
    \tau_{\rm IIB} =\tau_0 = \frac{\Sigma}{2} \quad\to\quad \tau_{\rm IIB}+m \,.
\end{equation}
This is precisely the parabolic monodromy of \cite{Hull:1998vy}.
Instead, the relative winding, $N=m_+-m_-$, measures the departure from the $Q$-invariant locus. Indeed, $p-\Sigma/2=-\Delta/2$ and \eqref{eq:marked_curve_winding} implies that $\Delta(y)=\Delta_0+Ny$. Together with \eqref{eq:0B_local_dictionary}, this shows that a path starting at $\Delta_0=0$ remains on the $Q$-invariant locus \eqref{eq:0B_symmetric_locus} when $N=0$, whereas $N\neq0$ necessarily produces a monodromy in the $Q$-odd sector. We will return to the dynamical consequences of this observation in the next subsection.

In section~\ref{sec:Hull}, we saw that $B(A,R)$ can also be viewed as an $S^1$ bundle over $T^2$. An analogous interpretation for massive type 0A provides a ``geometric'' picture of our M-theory construction: a doubling of the Hull construction, with two $S^1$ bundles over a common base and independent Chern numbers $m_\pm$. A subtlety arises because the junction cannot be chosen globally. As we explain below, the prescription of~\cite{Baykara:2026gem} resolves this apparent obstruction: the existence of the junction is physical, while its location is a gauge redundancy.

To illustrate our proposal, let $B_\pm$ be the $T^2_\pm$ bundles over $S^1_y$ for the two wedge factors, with complex structures $\tau_\pm(y)$. One can write a formal metric for each $B_\pm$,
\begin{equation}
    \dd s^2_{B_\pm} = R_y^2\dd y^2 +\frac{A_\pm}{{\rm Im}\tau_\pm(y)} \left| \dd z_\pm+\tau_\pm(y) \dd u\right|^2\,.
\end{equation}
Under $y\to y+1$, the shift $\tau_\pm\to\tau_\pm+m_\pm$ can be compensated by the large diffeomorphism
\begin{equation}
    z_\pm\to z_\pm-m_\pm u\,.
\end{equation}
This is the transition function that glues the $T_\pm^2$ fibers around $S_y^1$. Therefore, in analogy with the massive type IIA setup, each $B_\pm$ can be viewed as an $S^1_\pm$ bundle over the common base 
\begin{equation}
    T^2_{y,u}=S^1_y\times S^1_u\,,
\end{equation}
with fiber coordinate $z_\pm$ and base coordinates $(y,u)$. First, let us identify the two Chern numbers. The one-form
\begin{equation}
    \eta_\pm = \dd z_\pm+A_{{\rm KK},\pm}
\label{eq:branch_global_one_forms}
\end{equation}
is globally defined on $B_\pm$, although $ A_{{\rm KK},\pm}= m_\pm y\,\dd u$ is only a locally defined Kaluza--Klein connection on the base. Its curvature is given by
\begin{equation}
    F_{{\rm KK},\pm}=\dd A_{{\rm KK},\pm}=m_\pm\,\dd y\wedge \dd u \,.
\end{equation}
With respect to section~\ref{sec:Hull}, the structure is thus doubled: each circle bundle $B_\pm$ is characterized topologically by the first Chern class $ c_1(B_\pm) = \left[F_{{\rm KK},\pm}\right] \in H^2(T^2_{y,u},\mathbb Z)$, with
\begin{equation}
    \int_{T^2_{y,u}}c_1(B_\pm)=m_\pm \,.
\label{eq:branch_Chern_numbers}
\end{equation}
The parabolic monodromies $m_\pm$ of the two wedge factors are equivalent to the first Chern numbers, or geometric fluxes, of the two circle bundles $B_\pm$. 

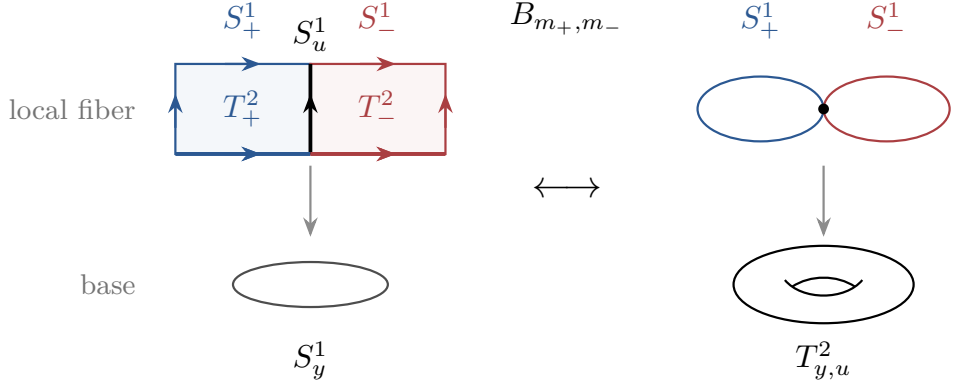
\begin{figure}[ht]
\centering
\resizebox{1\linewidth}{!}{%
\begin{tikzpicture}[
  font=\small,
  line width=.7pt,
  >=Stealth
]
\definecolor{plus}{RGB}{43,88,146}
\definecolor{minus}{RGB}{170,62,65}

\node at (2.85,3.15) {$B_{m_+,m_-}$};

\filldraw[draw=plus,fill=plus!5]
  (-1.5,1.65) rectangle (0,2.65);
\filldraw[draw=minus,fill=minus!5]
  (0,1.65) rectangle (1.5,2.65);

\draw[black,line width=1.2pt]
  (0,1.65)--(0,2.65);

\node[text=plus] at (-.75,2.15) {$T^2_+$};
\node[text=minus] at (.75,2.15) {$T^2_-$};

\node[above=4pt,text=plus] at (-.75,2.65) {$S^1_+$};
\node[above=0pt] at (0,2.65) {$S^1_u$};
\node[above=4pt,text=minus] at (.75,2.65) {$S^1_-$};

\foreach \x/\col in {-.95/plus,.55/minus}{
  \draw[->,\col] (\x,1.65)--(\x+.4,1.65);
  \draw[->,\col] (\x,2.65)--(\x+.4,2.65);
}

\foreach \x/\col in {-1.5/plus,0/black,1.5/minus}{
  \draw[->,\col] (\x,1.94)--(\x,2.32);
}

\draw[plus,line width=1.25pt]
  (-1.5,1.65)--(0,1.65);
\draw[minus,line width=1.25pt]
  (0,1.65)--(1.5,1.65);
\draw[->,plus] (-.95,1.65)--(-.55,1.65);
\draw[->,minus] (.55,1.65)--(.95,1.65);

\draw[->,black!45] (0,1.52)--(0,.73);
\draw[black!70] (0,.20) ellipse (.86 and .25);
\node[anchor=base] at (0,-.72) {$S^1_y$};

\node[font=\large] at (2.85,1.25) {$\longleftrightarrow$};

\draw[plus] (5.0,2.15) ellipse (.70 and .36);
\draw[minus] (6.4,2.15) ellipse (.70 and .36);
\fill (5.7,2.15) circle (1.7pt);

\node[above=4pt,text=plus] at (5.0,2.65) {$S^1_+$};
\node[above=4pt,text=minus] at (6.4,2.65) {$S^1_-$};

\draw[->,black!45] (5.7,1.52)--(5.7,.73);

\draw[black]
  (5.7,.20) ellipse (1 and .43);
\draw[black]
  (5.27,.25)
  .. controls (5.46,.02) and (5.94,.02) ..
  (6.13,.25);
\draw[black]
  (5.35,.175)
  .. controls (5.55,.31) and (5.85,.31) ..
  (6.05,.175);

\node[anchor=base] at (5.7,-.72) {$T^2_{y,u}$};

\node[
  text=black!55,
  font=\footnotesize,
  anchor=east
] at (-1.8,2.15) {local fiber};

\node[
  text=black!55,
  font=\footnotesize,
  anchor=east
] at (-1.8,.20) {base};

\pgfresetboundingbox
\path[use as bounding box]
  (-3.5,-1.0) rectangle (9.2,3.6);

\end{tikzpicture}%
}
\caption{Two local descriptions of the quantum fibration $B_{m_+,m_-}$: $T^2_\pm$ fibers joined along $S^1_u$ over $S^1_y$, and $S^1_\pm$ fibers joined over the common base $T^2_{y,u}$. The junction identifications are understood in the generalized sense described in the text.}
\label{fig:quantum-fibration}
\end{figure}

Crucially, we still have to glue $B_+$ and $B_-$ so that the full space is the same as the initial $T^2_+\vee_{S^1_u}T^2_-$ bundle over $S^1_y$. We denote this by
\begin{equation}
    B_{m_+,m_-} \coloneqq B_+\mathop{\vee}_{T^2_{y,u}} B_- \,,
\label{eq:quantum_fibration}
\end{equation}
and show its local structure in Figure~\ref{fig:quantum-fibration}. There is an apparent issue with reproducing the full space by gluing $B_+$ and $B_-$: at each point of the base $T^2_{y,u}$, we should join the two circle fibers to obtain $S^1_+\vee S^1_-$. However, the choice of a globally defined junction would imply the existence of global sections $s_\pm:T^2_{y,u}\to B_\pm$ for each bundle (to identify the junction point on both). 
When the Chern classes are not trivial, i.e., for non-vanishing values of $m_\pm$, this is not possible because no globally defined section exists. The resolution to this apparent puzzle comes from the interpretation of the junction point reviewed in section~\ref{sec:0AfromMtheory}; see the discussion around \eqref{eq:BDVequivalence}. In fact, the junction is ``delocalized'', and one does not need the sections $s_\pm$ to define the wedge sum. 

To see this explicitly, let $\{U_i\}$ be a cover of $S_y^1$, and choose a lift $y_i$ of the periodic coordinate on each patch. On the overlap of two patches, $U_i\cap U_j$, the two lifts are related by
\begin{equation}
    y_i=y_j+l_{ij} \,, \qquad l_{ij}\in\Z \,.
\end{equation}
The connection is
\begin{equation}
    A_{{\rm KK},\pm}^{(i)}=m_\pm y_i \, \dd u \,,
\end{equation}
and therefore, on the overlap,
\begin{equation}
    A_{{\rm KK},\pm}^{(i)}-A_{{\rm KK},\pm}^{(j)}=m_\pm l_{ij}\,\dd u=\dd (m_\pm l_{ij} u).
\end{equation}
Hence, the fiber coordinates must be relabeled by\footnote{This is a ${\rm U}(1)$-valued transition function since $m_\pm l_{ij}u$ shifts by an integer under $u\to u+1$.}
\begin{equation}
    z_{\pm,i}=z_{\pm,j}-m_\pm l_{ij}u\,,
\label{eq:branch_bundle_transition_functions}
\end{equation}
leaving the one-forms \eqref{eq:branch_global_one_forms} invariant.
One can now see the obstruction to a globally defined junction: on each patch, one may choose local representatives of the junction, $z_{\pm,i}=s_{\pm,i}(y_i,u)$. Then, on the overlap $U_i\cap U_j$, the point selected by $s_{\pm,j}$, when expressed in the $i$-th trivialization, has coordinate $s_{\pm,j}-m_\pm l_{ij}u \pmod{1}$. If these local representatives combined into global sections, they would have to satisfy
\begin{equation}
    s_{\pm,i} = s_{\pm,j}-m_\pm l_{ij}u \pmod{1}
\end{equation}
on each overlap. However, transporting a section once around $S^1_y$ would change its winding number along $S^1_u$ by $-m_\pm$, whereas continuity requires it to remain constant. Thus, no such compatible collection exists for non-zero $m_\pm$: the $s_{\pm,i}$ can only be retained as patchwise choices and not as local representatives of a globally defined junction section. This is not a problem if \eqref{eq:quantum_fibration} is to be understood in the generalized sense of the quantum compactification proposed in \cite{Baykara:2026gem}. 
As argued after \eqref{eq:branch_bundle_transition_functions}, two patchwise choices for the junction differ by a rotation of the fiber $S^1$ by $m_\pm l_{ij} u$, but such rotations are unphysical by \eqref{eq:BDVequivalence}. This is the new ingredient needed to define this \emph{quantum fibration} that yields massive type 0A. Explicitly, the $\mathrm{U}(1)_+\times\mathrm{U}(1)_-$ equivalence \eqref{eq:BDVequivalence} that moves the junction must here be imposed fiberwise in each local trivialization, with translation parameters allowed to depend on the base coordinates $(y,u)$:\footnote{This is natural since $\alpha$ and $\beta$ represent gauge symmetries and therefore can depend on the base coordinates.} 
\begin{equation}
    f_+(z_+) \sim f_+\left(z_+ + \alpha(y,u)\right) \, , \qquad f_-(z_-) \sim f_-\left(z_- \pm \beta(y,u)\right)\,,
\end{equation}
where $z_\pm$ are local coordinates along the circle fibers. As in \eqref{eq:BDVequivalence}, these conditions declare field configurations related by translations along the two circle fibers to be equivalent. Consequently, the local position of the junction is not physical: changing it can be compensated by translating the corresponding field configuration.
On the overlap $U_i\cap U_j$, the transition functions \eqref{eq:branch_bundle_transition_functions} correspond to the specific choice
\begin{equation}
    \alpha_{ij}=m_+ l_{ij}u \,, \qquad \pm\beta_{ij}=m_- l_{ij} u \,,
\end{equation}
where the sign is chosen consistently with the choice of orientation of SSP or SSP$'$. Under these transition maps, the local fields transform by pullback, and the junction representatives are compared only modulo independent circle rotations so that the values and derivatives of the fields along the circle fibers at the junction are independent of the chosen local trivialization. Hence, local representatives on different patches need only belong to the same equivalence orbit. In practice, the transition functions move the junction only by a gauge transformation and are therefore harmless in the definition of the quantum fibration.

Note that with this patchwise equivalence, DRP fields are glued independently on $B_+$ and $B_-$, while SSP and SSP$'$ fields are correlated at the local junction. Denoting its coordinate representatives on $U_i$ by $(s_{+,i},s_{-,i})$, the matching condition for local scalar fields $f_{\pm,i}$ is
\begin{equation}
    \left. \left(\frac{\d}{\d z_+}\right)^k f_{+,i} \right|_{s_{+,i}} = \epsilon^k \left.\left(\frac{\d}{\d z_-}\right)^k f_{-,i}\right|_{s_{-,i}}\,.
\label{eq:fibered_SSP}
\end{equation}
For $k=0$, this is the equality of the field values at the junction, while for $k>0$, it matches their derivatives along the circle fibers. The choices $\epsilon=+1$ and $\epsilon=-1$ correspond to SSP and SSP$'$. By the construction above, this matching condition is independent of the chosen local trivialization and of the representatives of the junction equivalence class.

The discussion above applies to scalar fields. Tensor fields are transported between local trivializations by the usual pullback under the transition diffeomorphisms, and their matching conditions are transformed accordingly. For the gravitino, one must additionally choose lifts of these diffeomorphisms to the spin bundles. We assume that the spin structures and their lifts are chosen so that the local SSP$^*$ condition,
\begin{equation}
    \left.\Psi_{+,i}\right|_{s_{+,i}}=-\left.\Psi_{-,i}\right|_{s_{-,i}}\,,
\end{equation}
is mapped to the same condition on every overlap. Under this assumption, the SSP$^*$ identification is patchwise consistent in the same way as the scalar matching conditions above.

With these prescriptions, the construction can accommodate arbitrary integral values of $(m_+,m_-)$ without enforcing either a global choice of the junction or a relation such as $m_+=m_-$. The quantum fibration is therefore not an ordinary fiberwise wedge; rather, it consists of the patchwise compactification data of the two bundles, the local field spaces with their resolution conditions, and the equivalences that relate them on the overlaps.

\subsection{Two Romans masses from M-theory}
\label{sec:mpmfromM}

The two integers $m_\pm$ characterizing the quantum fibration become the two Romans masses of type 0A after T-duality along $S^1_y$. Recalling that the unit-period type 0B axions are $C_0^\pm$, the ansatz \eqref{eq:tau_pm_winding} implies that
\begin{equation}
    C_0^\pm(x,y) = C_{0,0}^\pm(x)+m_\pm y \,.
\label{eq:C0_pm_winding}
\end{equation}
Consequently, their one-form field strengths are
\begin{equation}
    F_1^\pm=\dd C_{0,0}^\pm+m_\pm\,\dd y\,.
\end{equation}
T-duality acts separately on the two diagonal RR sectors and maps the components of $F_1^\pm$ along $S^1_y$ to the type 0A zero-form field strengths, hence
\begin{equation}
    F_0^\pm=m_\pm\,.
\label{eq:F0_pm}
\end{equation}
This is the (doubled) counterpart of the final T-duality step in section~\ref{sec:Hull}. Combining \eqref{eq:F0_pm} with \eqref{eq:branch_Chern_numbers}, we obtain the key identification
\begin{equation}
    F_0^\pm=m_\pm=\int_{T^2_{y,u}}c_1(B_\pm)\in\Z\,,
\end{equation}
showing that the two first Chern numbers of the bundles become the two integrally quantized Romans masses of type 0A. This identification is a central result of our construction: a geometrization in M-theory of the two Romans mass deformations of massive type 0A string theory. This geometrization is possible thanks to the quantum fibration. 

The same construction can also encompass the massive deformation of the RR field strengths. In the democratic formulation, the constant shifts of $C_0^\pm$ extend to the polyform \cite{Meessen:2001wk},
\begin{equation}
    C_B^\pm\to C_B^\pm+a_\pm e^{B_2} \,,
\end{equation}
where $a_\pm$ are constant shifts in the unit-period normalization. In a local patch for the $B_2$ field, suppressing the usual Kaluza--Klein covariantization, the generalized Scherk--Schwarz ansatz can therefore be written as
\begin{equation}
    C_B^\pm(x,y)=C_B^\pm(x)+m_\pm y\,e^{B_2(x)}\,,
\label{eq:type0B_polyform_winding}
\end{equation}
and its zero-form components reproduce \eqref{eq:C0_pm_winding}. T-duality along $S^1_y$ maps \eqref{eq:type0B_polyform_winding} to type 0A, where the field strengths become
\begin{equation}
    F_A^\pm=\left(\dd - H_3\wedge\right)C_A^\pm+m_\pm e^{B_2}\,,
\end{equation}
which are precisely the massive field strengths introduced in section~\ref{sec:type0st}.

In our construction, the total and relative windings also clarify the relation between the geometry of the fibration and the tachyon dynamics. In terms of the variables introduced in \eqref{eq:marked_curve_winding},
\begin{equation} 
    m_+^2-m_-^2 =(m_+ + m_-)  (m_+ - m_-)= MN\,.
\end{equation}
As we reviewed in section~\ref{sec:type0st}, a background with $\mathcal T=0$ and no RR fluxes other than $F_0^\pm$ requires \eqref{eq:T0=0}, which now reads $MN=0$. The solution to this equation has two branches.
For $N=0$, one has $m_+=m_-$ and the fibration remains on the $Q$-invariant locus $p=\Sigma/2$, thus reproducing the type IIB subsector. 
For $M=0$, instead, one has $m_+=-m_-$. The relative RR axion winds and the marked point moves away from the midpoint, while the zero-form contribution to the tachyon equation still vanishes. Departure from the $Q$-invariant locus therefore does not by itself force a non-zero tachyon. 
Only generic windings with $MN\neq0$ source the tachyon, requiring its profile to depart from $\mathcal T=0$ or the presence of additional RR fluxes. 

Interpreting the sources for the Romans masses is slightly subtler than in type IIA because electric-magnetic duality exchanges the two diagonal RR sectors. For the zero- and ten-form field strengths, the duality relation recalled in section~\ref{sec:type0st} reads
\begin{equation}
    F_{10}^\pm =-f_\mp(\mathcal T)\,*F_0^\mp \,.
\end{equation}
A ${\rm D}8^+$-brane, which couples electrically to $C_9^+$, is magnetically charged under $F_0^-$ and thus produces a jump in $m_-$. Similarly, a ${\rm D}8^-$-brane produces a jump in $m_+$:
\begin{eq}
    {\rm D}8^+ : \qquad & \Delta m_- \neq0 \,, \qquad \Delta m_+=0 \,, \\
    {\rm D}8^- : \qquad & \Delta m_+ \neq0 \,, \qquad \Delta m_-=0 \,.
\label{eq:D8_crossed_mass_jumps}
\end{eq}
In particular, an elementary source takes a background satisfying $m_+^2=m_-^2$ to one in which  $m_+^2\neq m_-^2$. Considering only $F_0^\pm$, this means that D8-branes cannot separate two regions in which the tachyon remains fixed at $\mathcal T=0$. 

As we have already emphasized for the massive type IIA case in section~\ref{sec:Hull}, our derivation of the two Romans masses from M-theory is only a local description, and additional dependence on spacetime coordinates is required to obtain a global solution. This should be the analog of the Polchinski--Witten setup~\cite{Polchinski:1995df}, with ${\rm D}8^\pm$-branes sourcing the Romans masses and, as we have just seen, the tachyon. Consistently with the rest of this work, we have only focused on the local description, and we leave a detailed analysis of global solutions for future work.

To summarize, one can obtain massive type 0A string theory from M-theory on the quantum fibration \eqref{eq:quantum_fibration}, keeping the integers $m_\pm$ fixed while taking the type 0B zero-area limit \eqref{eq:0B_zero_area}. This gives type 0B compactified on $S^1_y$ with the Scherk--Schwarz ansatz \eqref{eq:tau_pm_winding}, and a final T-duality along $S^1_y$ leads to massive type 0A on the dual circle, which decompactifies as $R_y\to0$. 

\section{On branes and Chern--Simons couplings}
\label{sec:branes-CS}

Our discussion of RR fields in section~\ref{sec:mpmfromM} gives us a good opportunity to point out a puzzle regarding the type 0A branes.\footnote{A complementary analysis of type 0A brane charges during the collapse of one wedge factor has been given in \cite{Kamal:2026msr}.}
This has some consequences for the Chern--Simons couplings in both  massless and massive type 0A, which we discuss at the end of the section.

The charged branes of type 0A string theory couple to the tachyon, and branes of the $\pm$ types have opposite linear couplings. At leading order in $\mathcal{T}$, D$p^\pm$-branes have tension proportional to $1\mp\mathcal{T}/4$, in our conventions~\cite{Garousi:1999fu}. Based on the low-order expansion in the tachyon expectation value, it has also been conjectured in~\cite{Garousi:1999fu} that the full closed-form function reads
\begin{equation}\label{eq:tension_conjecture}
    T_{{\rm D}p^{\pm}}
    =\frac{M_S^{p+1}}{\sqrt{2}(2\pi)^p g_s}
    \left(1\pm\frac{\mathcal{T}}{2}\right)^{-\frac{1}{2}}\,.
\end{equation}
On the other hand, the radii of the two wedge factors depend on the tachyon as~\cite{Baykara:2026gem}
\begin{equation}\label{eq:Rpm}
    R_\pm=R_0\left(1\pm\frac{\mathcal{T}}{2}\right)^{\frac{1}{2}}\,,
    \qquad \text{where}\qquad
    R_0=\frac{\sqrt{2}g_s}{M_S}\,.
\end{equation}
The DRP nature of the RR fields suggests that in M-theory their sources should arise from the two wedge factors separately. Indeed, as discussed in~\cite{Baykara:2026gem}, the two types of D0-branes can be interpreted as KK momenta on $S^1_\pm$, with mass
\begin{equation}\label{eq:D0_mass}
    m_{{\rm D}0^\pm}\propto \left(1\pm \frac{\mathcal{T}}{2}\right)^{-\frac{1}{2}} \sim \frac{1}{R_\pm}\,.
\end{equation}
One might hope to extend this reasoning to the other type 0A branes. However, this extension is not straightforward, and the resulting picture is more involved.

Consider first D$4^\pm$-branes. Their tension is again given by \eqref{eq:tension_conjecture}, and their M-theory origin must be M5-branes wrapped on the two wedge factors. However, an M5-brane wrapped on $S^1_\pm$ has tension proportional to $R_\pm$, which does not match that of a D$4^\pm$-brane. Instead, it matches the tension of a D$4^\mp$-brane, but only at linear order in $\mathcal{T}$; the agreement fails at higher orders.

A similar analysis encounters further difficulties for D2- and D6-branes. Indeed, consider the ratio of the tensions of type 0A D$p^+$- and D$p^-$-branes, assuming the conjecture of~\cite{Garousi:1999fu}: 
\begin{equation}\label{eq:rho_garousi}
    \rho_p=\frac{T_{{\rm D}p^+}}{T_{{\rm D}p^-}}=\sqrt{\frac{1-\frac{\mathcal{T}}{2}}{1+\frac{\mathcal{T}}{2}}}=\frac{R_-}{R_+}\,,
\end{equation}
where we used \eqref{eq:Rpm}. If the corresponding M-theory objects have tensions proportional to $R_\pm^n$, assigning D$p^+$ and D$p^-$ to $S^1_+$ and $S^1_-$ respectively, or reversing this assignment, gives
\begin{equation}
    \rho_M=\left(\frac{R_-}{R_+}\right)^{\pm n}\,.
\label{eq:tension_ration_D+D-}
\end{equation}
Only objects with $n=\pm1$ can therefore reproduce \eqref{eq:rho_garousi}. Among the standard M-theory objects, these are KK momentum modes, with $n=-1$, and wrapped branes, with $n=1$. This is consistent with the successful matching of D$0^\pm$ masses and with the matching of D$4^\pm$ tensions at linear order. Besides, the tension of fundamental type 0A strings arises from M2-branes wrapping the connected resolution, as discussed in~\cite{Baykara:2026gem}.

In particular, the tensions of type 0A D2-branes cannot be reproduced by the ``classical'' expression for unwrapped M2-branes in M-theory. This is a puzzle of the formulation, but perhaps not entirely surprising: the branes whose tensions are reproduced, at least at leading order, are either wrapped or associated with quantized momentum. These two setups can remain meaningful when the internal space is not geometrical and is instead interpreted through its space of functions (recall our discussion in section~\ref{sec:0AfromMtheory}). The cases left out, D2- and D6-branes (with $n=0$ and $n=2$ in \eqref{eq:tension_ration_D+D-}, respectively), are those for which the classical notion of an internal M-theory circle is most critical: D2-branes require \emph{localizing} degrees of freedom on the M-theory circles, while D6-branes are \emph{pure geometry} in M-theory, which is difficult to reconcile with the quantum-geometric formulation of the wedge sum.
Note that we are not claiming that D$2^\pm$-branes do not come from M2-branes in M-theory; rather, we are arguing that the classical tension formula need not apply unchanged when the M2-branes are to be \emph{localized} on $S^1_+\vee S^1_-$. Indeed, one could envisage that the wavefunction of a ``localized'' M2-brane spreads over the wedge factor on which it ``localizes'', and this could potentially reproduce the tachyon-dependence of the tension.

We currently have no concrete proposal for resolving this puzzle, but would like to mention a related observation. In the democratic formulation of type 0A string theory, one takes RR field strengths of only one type, $+$ or $-$, with no self-duality constraint among them. Indeed, the electric-magnetic dual of a form of type $\pm$ belongs to type $\mp$; for example,
\begin{equation}
    F_8^\mp=f_\pm(\mathcal{T})\,*F_2^\pm\,,
    \qquad
    F_6^\mp=-f_\pm(\mathcal{T})\,*F_4^\pm\,.
\label{eq:crossed_duality_D0_D2}
\end{equation}
This has an interesting implication for the D-branes: given D$0^\pm$- and D$4^\pm$-branes, one obtains D$6^\mp$- and D$2^\mp$-branes by electric-magnetic duality. Thus, the four species D$0^\pm$ and D$4^\pm$ represent all the D0-, D2-, D4-, and D6-branes up to electric-magnetic duality. Since the direct M-theory interpretation of D2- and D6-branes relies on a classical internal geometry, one can instead focus on D0- and D4-branes, whose description is more directly compatible with the quantum geometry of~\cite{Baykara:2026gem}.

Section~\ref{sec:m0AfromMtheory} allows us to include ${\rm D}8^\pm$-branes in this discussion. According to \eqref{eq:D8_crossed_mass_jumps}, a ${\rm D}8^\pm$-brane is a domain wall across which the Chern number $m_\mp$ of the opposite wedge-factor bundle jumps. Thus, as for ${\rm D}4^\pm$-branes, a brane of one type is associated with the wedge factor of the other type.

The discussion above revolves around the exchange of brane types under the duality relations \eqref{eq:crossed_duality_D0_D2}. In the $S^1_+\vee S^1_-$ picture, this suggests a prescription correlating the two wedge factors, possibly through the junction.
A related observation concerns the Chern--Simons terms. In the democratic formulation of type 0A, as in type IIA, there is no explicit Chern--Simons term in the action; such terms appear upon dualizing the higher-degree RR field strengths. In the massless type 0A theory, the Chern--Simons term is \cite{Meessen:2001wk}
\begin{equation}
    S_{\rm CS}^{(0A)} \ \sim \ \int B_2\wedge \dd C_3^+\wedge \dd C_3^- \,,
\label{eq:type0A_mixed_CS}
\end{equation}
which can also be expressed as the difference between two type IIA-like Chern--Simons terms for $C_3$ and $C_3'$. While its type IIA counterpart follows from the M-theory
Chern--Simons term,
\begin{equation}
    S_{\rm CS}^{(11)}\ \sim \ \int \mathcal C_3\wedge\mathcal G_4\wedge\mathcal G_4 \,,
\end{equation}
deriving \eqref{eq:type0A_mixed_CS} from the quantum compactification is less straightforward. In fact, the ansatz
\begin{equation}
    \left.\mathcal C_3\right|_{S^1_\pm} = C_3^\pm+B_2\wedge \dd\theta_\pm
\end{equation}
would yield only diagonal contributions $B_2\wedge \dd C_3^\pm\wedge \dd C_3^\pm$, rather than the mixed type 0A coupling. This suggests that additional data correlating the two wedge factors are needed, possibly through the junction prescription. Analogously, in massive type 0A, the Chern--Simons term contains the additional contributions~\cite{Meessen:2001wk}
\begin{equation}
    \int \left(\frac{1}{6}F_0^-\,\dd C_3^+\wedge B_2^3+\frac{1}{6}F_0^+\,\dd C_3^-\wedge B_2^3+\frac{1}{20}F_0^+F_0^-\,B_2^5\right) \,,
\label{eq:massive_type0A_CS}
\end{equation}
where $B_2^k$ denotes the $k$-fold wedge product. The same exchange of RR labels appears here: $F_0^\mp$ pairs with $C_3^\pm$, consistent with ${\rm D}8^\pm$-branes sourcing $m_\mp$. In our dictionary, these couplings again correlate data associated with opposite wedge factors.

Deriving the massless and massive Chern--Simons terms of type 0A from M-theory remains an open problem. Our discussion suggests that the junction may play a key role, together with the electric-magnetic duality relating the two RR sectors. Understanding this structure may also help resolve the puzzles concerning D2- and D6-brane tensions. 

\section{Conclusions}\label{sec:conclusions}

In this work, building on the recent proposal of~\cite{Baykara:2026gem}, we proposed an M-theory origin for the two massive deformations of type 0A string theory. Our construction extends Hull's M-theory description of massive type IIA~\cite{Hull:1998vy} to the non-supersymmetric setting of type 0A, where the doubled RR sector and the field-dependent compactification prescription require additional structure. In particular, the two Romans masses arise through independent duality twists in type 0B, despite the presence of only a single ${\rm SL}(2,\Z)$ duality factor. We argued that both twists can be encoded in an elliptic curve with a marked point, providing a geometric description of the type 0B duality group proposed in~\cite{Baykara:2026gem}. Our main contribution is the \emph{quantum fibration} of section~\ref{ssec:quantum_fibration}, whose two wedge-factor bundles carry independent Chern numbers that become the two Romans masses after T-duality. We thus argued that M-theory on this space is dual to massive type 0A string theory.

Our quantum fibration extends the junction prescription of the $S^1_+\vee S^1_-$ compactification to a setting with non-trivial bundle topology. This suggests that the same approach may apply more broadly, opening up a wider class of M-theory compactifications on higher-dimensional internal spaces with analogous singularities. A natural direction would be to consider $T^p$ bundles over $T^q_+\vee_{T^r}T^q_-$, where the two base tori are glued along a common $T^r$ with $r<q$, and investigate how the field-dependent resolution conditions extend to these spaces.

Other deformations of (super)gravity theories could also be considered. Classifying the deformations of the type 0 low-energy actions and investigating their origin in quantum compactifications of M-theory would provide an interesting test of the duality conjectured in~\cite{Baykara:2026gem}. For instance, one could extend our construction using a non-parabolic element of the type 0B duality group.
The key feature that allowed us to obtain massive type 0A, and could make this broader program feasible, is that many deformations arise from Scherk--Schwarz reductions. Since these can be formulated through boundary conditions on fields, they fit naturally within the prescription of~\cite{Baykara:2026gem}, which focuses on the space of functions rather than on the underlying topological space.

Our observations in section~\ref{sec:branes-CS} on the crossed structure of type 0 RR fields and branes raise puzzles about the M-theory origin of branes and Chern--Simons terms. We have seen that issues arise when forcing a geometric interpretation of the internal $S^1_+\vee S^1_-$. This is perhaps not surprising, but if the interpretation of type 0A as M-theory on this space is to be meaningful, there must be a way to account for the missing branes and the Chern--Simons terms. These are not the only subtleties with branes in the quantum compactification: for instance, \cite{Altavista:2026evd} and \cite{Baykara:2026vdc} gave two different proposals for type 0 D9-branes. For the charged branes of type 0A and its massive deformations, we have argued that correlations between the wedge factors, possibly through the junction, should play a key role.

A broader question concerns the dynamics. The construction of \cite{Baykara:2026gem} is presently formulated mainly at the level of field assignments and junction conditions, and our extension necessarily inherits this limitation. It would be important to find a more fundamental description of the quantum fibration determining how the field space arises from M-theory.

While these questions remain open, our findings provide a further test of the proposal of~\cite{Baykara:2026gem} through the matching of the quantized Romans masses. It would be interesting to identify other discrete or topological quantities to match between non-supersymmetric string theories and quantum compactifications of M-theory.\footnote{A recent program along these lines investigates the global topology of gauge groups in non-supersymmetric string theories; see, e.g.,~\cite{Fraiman:2023cpa,Larotonda:2024thv,Fraiman:2026ltu}.} Such matches would provide non-trivial tests of string dualities in the absence of spacetime supersymmetry.

\paragraph{Acknowledgments}
We would like to thank Emilian Dudas for helpful comments on the manuscript.
The work of N.C.~is supported by the Research Foundation Flanders (FWO grant 1259125N).
A.M.~is supported by a Juan de la Cierva contract (JDC2024-054820-I) from Spain’s Ministry of Science, Innovation and Universities.
S.R.~is supported by the ERC Starting Grant QGuide101042568 - StG 2021.
This publication has been funded within the framework of the R\&D\&I Projects PID2024-156043NB-I00 and CEX2025-001574-S, funded by MICIU\slash AEI\slash10.13039\slash501100011033. The research presented in this publication falls within the research line Strings and Quantum Gravity.
This work was performed in part at the Aspen Center for Physics, which is supported by a grant from the Simons Foundation (1161654, Troyer), and was supported by two short term scientific mission grants from the COST action CA22113 THEORY-CHALLENGES.

\paragraph*{AI assistance.}
Large language models were used to assist with editing and checking calculations in this manuscript. The authors take full responsibility for the content, accuracy, and conclusions of the work.

\bibliographystyle{utphys}
\bibliography{mybib}

@article{Hull:1998vy,
    author = "Hull, C. M.",
    title = "{Massive string theories from M theory and F theory}",
    eprint = "hep-th/9811021",
    archivePrefix = "arXiv",
    reportNumber = "QMW-PH-98-36, LPTENS-98-32",
    doi = "10.1088/1126-6708/1998/11/027",
    journal = "JHEP",
    volume = "11",
    pages = "027",
    year = "1998"
}

@article{Baykara:2026gem,
    author = "Baykara, Zihni Kaan and Dudas, Emilian and Vafa, Cumrun",
    title = "{M-theory on $S^1\vee S^1$ as Type 0A}",
    eprint = "2603.13468",
    archivePrefix = "arXiv",
    primaryClass = "hep-th",
    month = "3",
    year = "2026"
}

@article{Altavista:2026evd,
    author = "Altavista, Chiara and Anastasi, Edoardo and Raucci, Salvatore and Uranga, Angel M. and Wang, Chuying",
    title = "{Ho{\v{r}}ava-Witten theory on $S^1 \vee S^1$ as type 0 orientifold}",
    eprint = "2603.25786",
    archivePrefix = "arXiv",
    primaryClass = "hep-th",
    reportNumber = "IFT-UAM/CSIC-26-39",
    doi = "10.1103/b6kr-l2l2",
    journal = "Phys. Rev. D",
    volume = "114",
    number = "6",
    pages = "066008",
    year = "2026"
}

@article{Baykara:2026vdc,
    author = "Baykara, Zihni Kaan and Delgado, Matilda and Dudas, Emilian and De Freitas, Hector Parra and Vafa, Cumrun",
    title = "{A Duality Web for Non-Supersymmetric Strings}",
    eprint = "2604.07433",
    archivePrefix = "arXiv",
    primaryClass = "hep-th",
    month = "4",
    year = "2026"
}

@article{Altavista:2026brr,
    author = "Altavista, Chiara and Raucci, Salvatore and Uranga, Angel M. and Wang, Chuying",
    title = "{Heterotic ouroboros}",
    eprint = "2604.22915",
    archivePrefix = "arXiv",
    primaryClass = "hep-th",
    reportNumber = "IFT-UAM/CSIC-26-53",
    doi = "10.1007/JHEP07(2026)145",
    journal = "JHEP",
    volume = "07",
    pages = "145",
    year = "2026"
}

@article{Dasgupta:2026maq,
    author = "Dasgupta, Keshav and Tatar, Radu",
    title = "{Towards Wedge Construction of Four-Dimensional Non-Supersymmetric Theories and Torsion Classes}",
    eprint = "2605.05333",
    archivePrefix = "arXiv",
    primaryClass = "hep-th",
    month = "5",
    year = "2026"
}

@article{Basile:2026trt,
    author = {Basile, Ivano and L{\"u}st, Dieter},
    title = "{String dualities and wedge singularities}",
    eprint = "2606.05287",
    archivePrefix = "arXiv",
    primaryClass = "hep-th",
    reportNumber = "MPP-2026-102",
    month = "6",
    year = "2026"
}

@article{Meessen:2001wk,
    author = "Meessen, Patrick and Ortin, Tomas",
    title = "{Type 0 T duality and the tachyon coupling}",
    eprint = "hep-th/0103244",
    archivePrefix = "arXiv",
    reportNumber = "IFT-UAM-CSIC-01-07, KUL-TF-2001-9",
    doi = "10.1103/PhysRevD.64.126005",
    journal = "Phys. Rev. D",
    volume = "64",
    pages = "126005",
    year = "2001"
}

@article{Diaconescu:2000wy,
    author = "Diaconescu, Duiliu-Emanuel and Moore, Gregory W. and Witten, Edward",
    title = "{E(8) gauge theory, and a derivation of K theory from M theory}",
    eprint = "hep-th/0005090",
    archivePrefix = "arXiv",
    reportNumber = "IASSNS-HEP-00-39",
    doi = "10.4310/ATMP.2002.v6.n6.a2",
    journal = "Adv. Theor. Math. Phys.",
    volume = "6",
    pages = "1031--1134",
    year = "2003"
}

@article{Leone:2025mwo,
    author = "Leone, Giorgio and Raucci, Salvatore",
    title = "{Aspects of strings without spacetime supersymmetry}",
    eprint = "2509.24703",
    archivePrefix = "arXiv",
    primaryClass = "hep-th",
    reportNumber = "IFT-UAM/CSIC-25-100",
    doi = "10.1007/s40766-025-00078-z",
    journal = "Riv. Nuovo Cim.",
    volume = "49",
    number = "3",
    pages = "75--136",
    year = "2026"
}

@article{Romans:1985tz,
    author = "Romans, L. J.",
    editor = "Salam, A. and Sezgin, E.",
    title = "{Massive N=2a Supergravity in Ten-Dimensions}",
    reportNumber = "NSF-ITP-85-148",
    doi = "10.1016/0370-2693(86)90375-8",
    journal = "Phys. Lett. B",
    volume = "169",
    pages = "374",
    year = "1986"
}

@article{Bautier:1997yp,
    author = "Bautier, K. and Deser, Stanley and Henneaux, M. and Seminara, D.",
    title = "{No cosmological D = 11 supergravity}",
    eprint = "hep-th/9704131",
    archivePrefix = "arXiv",
    reportNumber = "ULB-TH-97-07, BRX-TH-411",
    doi = "10.1016/S0370-2693(97)00639-4",
    journal = "Phys. Lett. B",
    volume = "406",
    pages = "49--53",
    year = "1997"
}

@article{Howe:1997qt,
    author = "Howe, Paul S. and Lambert, N. D. and West, Peter C.",
    title = "{A New massive type IIA supergravity from compactification}",
    eprint = "hep-th/9707139",
    archivePrefix = "arXiv",
    reportNumber = "KCL-TH-97-46",
    doi = "10.1016/S0370-2693(97)01199-4",
    journal = "Phys. Lett. B",
    volume = "416",
    pages = "303--308",
    year = "1998"
}

@article{Witten:1995ex,
    author = "Witten, Edward",
    title = "{String theory dynamics in various dimensions}",
    eprint = "hep-th/9503124",
    archivePrefix = "arXiv",
    reportNumber = "IASSNS-HEP-95-18",
    doi = "10.1016/0550-3213(95)00158-O",
    journal = "Nucl. Phys. B",
    volume = "443",
    pages = "85--126",
    year = "1995"
}

@article{Townsend:1995kk,
    author = "Townsend, P. K.",
    title = "{The eleven-dimensional supermembrane revisited}",
    eprint = "hep-th/9501068",
    archivePrefix = "arXiv",
    reportNumber = "DAMTP-R-95-2",
    doi = "10.1016/0370-2693(95)00397-4",
    journal = "Phys. Lett. B",
    volume = "350",
    pages = "184--187",
    year = "1995"
}

@article{Lust:2017aqj,
    author = "Lust, Severin and Ruter, Philipp and Louis, Jan",
    title = "{Maximally Supersymmetric AdS Solutions and their Moduli Spaces}",
    eprint = "1711.06180",
    archivePrefix = "arXiv",
    primaryClass = "hep-th",
    reportNumber = "ZMP-HH-17-27, CPHT-RR054.102017, EMPG-17-20",
    doi = "10.1007/JHEP03(2018)019",
    journal = "JHEP",
    volume = "03",
    pages = "019",
    year = "2018"
}

@article{Bergshoeff:1997ak,
    author = "Bergshoeff, Eric and Lozano, Yolanda and Ortin, Tomas",
    title = "{Massive branes}",
    eprint = "hep-th/9712115",
    archivePrefix = "arXiv",
    reportNumber = "UG-8-97, QMW-PH-97-28, CERN-TH-97-229, IFT-UAM-CSIC-97-2",
    doi = "10.1016/S0550-3213(98)00045-5",
    journal = "Nucl. Phys. B",
    volume = "518",
    pages = "363--423",
    year = "1998"
}

@article{Farakos:2017jme,
    author = "Farakos, Fotis and Lanza, Stefano and Martucci, Luca and Sorokin, Dmitri",
    title = "{Three-forms in Supergravity and Flux Compactifications}",
    eprint = "1706.09422",
    archivePrefix = "arXiv",
    primaryClass = "hep-th",
    doi = "10.1140/epjc/s10052-017-5185-y",
    journal = "Eur. Phys. J. C",
    volume = "77",
    number = "9",
    pages = "602",
    year = "2017"
}

@article{Scherk:1979zr,
    author = "Scherk, Joel and Schwarz, John H.",
    editor = "Salam, A. and Sezgin, E.",
    title = "{How to Get Masses from Extra Dimensions}",
    reportNumber = "LPTENS-79-2",
    doi = "10.1016/0550-3213(79)90592-3",
    journal = "Nucl. Phys. B",
    volume = "153",
    pages = "61--88",
    year = "1979"
}

@article{Bergshoeff:1996ui,
    author = "Bergshoeff, E. and de Roo, M. and Green, Michael B. and Papadopoulos, G. and Townsend, P. K.",
    title = "{Duality of type II 7 branes and 8 branes}",
    eprint = "hep-th/9601150",
    archivePrefix = "arXiv",
    reportNumber = "DAMTP-R-95-55-REV, UG-15-95",
    doi = "10.1016/0550-3213(96)00171-X",
    journal = "Nucl. Phys. B",
    volume = "470",
    pages = "113--135",
    year = "1996"
}

@article{Tachikawa:2018njr,
    author = "Tachikawa, Yuji and Yonekura, Kazuya",
    title = "{Why are fractional charges of orientifolds compatible with Dirac quantization?}",
    eprint = "1805.02772",
    archivePrefix = "arXiv",
    primaryClass = "hep-th",
    reportNumber = "IPMU-18-0067",
    doi = "10.21468/SciPostPhys.7.5.058",
    journal = "SciPost Phys.",
    volume = "7",
    number = "5",
    pages = "058",
    year = "2019"
}

@article{Kamal:2026msr,
    author = "Kamal, Ahmed Rakin",
    title = "{A Circle That Won't Return: The Fate of RR Fluxes and D-branes in Type 0A Tachyon Condensation}",
    eprint = "2606.26280",
    archivePrefix = "arXiv",
    primaryClass = "hep-th",
    month = "6",
    year = "2026"
}

@article{Lavrinenko:1997qa,
    author = "Lavrinenko, I. V. and Lu, Hong and Pope, C. N.",
    title = "{Fiber bundles and generalized dimensional reduction}",
    eprint = "hep-th/9710243",
    archivePrefix = "arXiv",
    reportNumber = "CTP-TAMU-43-97, LPTENS-97-50",
    doi = "10.1088/0264-9381/15/8/008",
    journal = "Class. Quant. Grav.",
    volume = "15",
    pages = "2239--2256",
    year = "1998"
}

@article{Robbins:2021ibx,
    author = "Robbins, Daniel G. and Sharpe, Eric and Vandermeulen, Thomas",
    title = "{Quantum symmetries in orbifolds and decomposition}",
    eprint = "2107.12386",
    archivePrefix = "arXiv",
    primaryClass = "hep-th",
    doi = "10.1007/JHEP02(2022)108",
    journal = "JHEP",
    volume = "02",
    pages = "108",
    year = "2022"
}

@article{Kaidi:2019tyf,
    author = "Kaidi, Justin and Parra-Martinez, Julio and Tachikawa, Yuji",
    title = "{Topological Superconductors on Superstring Worldsheets}",
    eprint = "1911.11780",
    archivePrefix = "arXiv",
    primaryClass = "hep-th",
    reportNumber = "IPMU-19-0164, UCLA/TEP/2019/106",
    doi = "10.21468/SciPostPhys.9.1.010",
    journal = "SciPost Phys.",
    volume = "9",
    pages = "10",
    year = "2020"
}

@article{Klebanov:1998yya,
    author = "Klebanov, Igor R. and Tseytlin, Arkady A.",
    title = "{D-branes and dual gauge theories in type 0 strings}",
    eprint = "hep-th/9811035",
    archivePrefix = "arXiv",
    reportNumber = "PUPT-1819, IMPERIAL-TP-98-99-07",
    doi = "10.1016/S0550-3213(99)00041-3",
    journal = "Nucl. Phys. B",
    volume = "546",
    pages = "155--181",
    year = "1999"
}

@article{Bergman:1999km,
    author = "Bergman, Oren and Gaberdiel, Matthias R.",
    title = "{Dualities of type 0 strings}",
    eprint = "hep-th/9906055",
    archivePrefix = "arXiv",
    reportNumber = "CALT-68-2228, DAMTP-1999-74",
    doi = "10.1088/1126-6708/1999/07/022",
    journal = "JHEP",
    volume = "07",
    pages = "022",
    year = "1999"
}

@article{Costa:2000nw,
    author = "Costa, Miguel S. and Gutperle, Michael",
    title = "{The Kaluza-Klein Melvin solution in M theory}",
    eprint = "hep-th/0012072",
    archivePrefix = "arXiv",
    reportNumber = "LPTENS-00-44, HUTP-00-A048",
    doi = "10.1088/1126-6708/2001/03/027",
    journal = "JHEP",
    volume = "03",
    pages = "027",
    year = "2001"
}

@article{Russo:2001tf,
    author = "Russo, J. G. and Tseytlin, Arkady A.",
    title = "{Magnetic backgrounds and tachyonic instabilities in closed superstring theory and M theory}",
    eprint = "hep-th/0104238",
    archivePrefix = "arXiv",
    reportNumber = "OHSTPY-HEP-T-01-009",
    doi = "10.1016/S0550-3213(01)00358-3",
    journal = "Nucl. Phys. B",
    volume = "611",
    pages = "93--124",
    year = "2001"
}

@article{Garousi:1999fu,
    author = "Garousi, Mohammad R.",
    title = "{String scattering from D-branes in type 0 theories}",
    eprint = "hep-th/9901085",
    archivePrefix = "arXiv",
    reportNumber = "IPM-P-99-5",
    doi = "10.1016/S0550-3213(99)00152-2",
    journal = "Nucl. Phys. B",
    volume = "550",
    pages = "225--237",
    year = "1999"
}

@article{Moore:2002cp,
    author = "Moore, Gregory W. and Saulina, Natalia",
    title = "{T duality, and the K theoretic partition function of type IIA superstring theory}",
    eprint = "hep-th/0206092",
    archivePrefix = "arXiv",
    reportNumber = "RUNHETC-2002-15, NI-02013-MTH",
    doi = "10.1016/j.nuclphysb.2003.07.028",
    journal = "Nucl. Phys. B",
    volume = "670",
    pages = "27--89",
    year = "2003"
}

@misc{Sagnotti:1982ez,
    author = "Sagnotti, Augusto and Tomaras, Theodore N.",
    title = "PROPERTIES OF ELEVEN-DIMENSIONAL SUPERGRAVITY",
    howpublished = "CALT-68-885",
    year = "1982"
}

@inproceedings{Deser:1997gm,
    author = "Deser, Stanley",
    title = "{Uniqueness of D = 11 supergravity}",
    booktitle = "{6th Conference on Quantum Mechanics of Fundamental Systems: Black Holes and the Structure of the Universe}",
    eprint = "hep-th/9712064",
    archivePrefix = "arXiv",
    reportNumber = "ULB-TH-97-07A, BRX-TH-424",
    pages = "70--80",
    month = "12",
    year = "1997"
}

@inproceedings{Dabholkar:1997zd,
    author = "Dabholkar, Atish",
    title = "{Lectures on orientifolds and duality}",
    booktitle = "{ICTP Summer School in High-Energy Physics and Cosmology}",
    eprint = "hep-th/9804208",
    archivePrefix = "arXiv",
    reportNumber = "TIFR-TH-98-13",
    pages = "128--191",
    month = "6",
    year = "1997"
}

@article{Pantev:2016nze,
    author = "Pantev, T. and Sharpe, E.",
    title = "{Duality group actions on fermions}",
    eprint = "1609.00011",
    archivePrefix = "arXiv",
    primaryClass = "hep-th",
    doi = "10.1007/JHEP11(2016)171",
    journal = "JHEP",
    volume = "11",
    pages = "171",
    year = "2016"
}

@article{Dudas:2001wd,
    author = "Dudas, E. and Mourad, J. and Sagnotti, A.",
    title = "{Charged and uncharged D-branes in various string theories}",
    eprint = "hep-th/0107081",
    archivePrefix = "arXiv",
    reportNumber = "LPT-ORSAY-01-56, ROM2F-01-18",
    doi = "10.1016/S0550-3213(01)00552-1",
    journal = "Nucl. Phys. B",
    volume = "620",
    pages = "109--151",
    year = "2002"
}

@inproceedings{Ginsparg:1988ui,
    author = "Ginsparg, Paul H.",
    title = "APPLIED CONFORMAL FIELD THEORY",
    booktitle = "{Les Houches Summer School in Theoretical Physics: Fields, Strings, Critical Phenomena}",
    eprint = "hep-th/9108028",
    archivePrefix = "arXiv",
    reportNumber = "HUTP-88-A054",
    month = "9",
    year = "1988"
}

@article{Kan:2026sea,
    author = "Kan, Naoto and Kawahira, Masashi and Wada, Hiroki",
    title = "{7-branes and $\Gamma_0(2)$ Duality in Type 0B String Theory}",
    eprint = "2608.16522",
    archivePrefix = "arXiv",
    primaryClass = "hep-th",
    reportNumber = "OU-HET-1321, YITP-26-109, TU-1316",
    month = "8",
    year = "2026"
}

@article{Mourad:2017rrl,
    author = "Mourad, J. and Sagnotti, A.",
    title = "{An Update on Brane Supersymmetry Breaking}",
    eprint = "1711.11494",
    archivePrefix = "arXiv",
    primaryClass = "hep-th",
    month = "11",
    year = "2017"
}

@article{Basile:2021vxh,
    author = "Basile, Ivano",
    title = "{Supersymmetry breaking and stability in string vacua: Brane dynamics, bubbles and the swampland}",
    eprint = "2107.02814",
    archivePrefix = "arXiv",
    primaryClass = "hep-th",
    doi = "10.1007/s40766-021-00024-9",
    journal = "Riv. Nuovo Cim.",
    volume = "44",
    number = "10",
    pages = "499--596",
    year = "2021"
}

@book{Angelantonj:2024tns,
    author = "Angelantonj, Carlo and Florakis, Ioannis",
    title = "{A Lightning Introduction to String Theory}",
    eprint = "2406.09508",
    archivePrefix = "arXiv",
    primaryClass = "hep-th",
    doi = "10.1007/978-981-19-3079-9_53-1",
    year = "2024",
    publisher = "Handbook of Quantum Gravity"
}

@phdthesis{Raucci:2024fnp,
    author = "Raucci, Salvatore",
    title = "{Spacetime aspects of non-supersymmetric strings}",
    eprint = "2409.19395",
    archivePrefix = "arXiv",
    primaryClass = "hep-th",
    school = "Pisa, Scuola Normale Superiore",
    month = "9",
    year = "2024"
}

@article{Dudas:2025ubq,
    author = "Dudas, E. and Mourad, J. and Sagnotti, A.",
    title = "{Supersymmetry breaking with fields, strings and branes}",
    eprint = "2511.04367",
    archivePrefix = "arXiv",
    primaryClass = "hep-th",
    doi = "10.1016/j.physrep.2026.02.005",
    journal = "Phys. Rept.",
    volume = "1175",
    pages = "1--256",
    year = "2026"
}

@article{Aharony:2010af,
    author = "Aharony, Ofer and Jafferis, Daniel and Tomasiello, Alessandro and Zaffaroni, Alberto",
    title = "{Massive type IIA string theory cannot be strongly coupled}",
    eprint = "1007.2451",
    archivePrefix = "arXiv",
    primaryClass = "hep-th",
    doi = "10.1007/JHEP11(2010)047",
    journal = "JHEP",
    volume = "11",
    pages = "047",
    year = "2010"
}

@article{Martelli:geomtop,
    author = "Martelli, Bruno",
    title = "{An Introduction to Geometric Topology}",
    eprint = "1610.02592",
    archivePrefix = "arXiv",
    primaryClass = "math.GT",
    year = "2022"
}

@article{Fraiman:2023cpa,
    author = "Fraiman, Bernardo and Gra{\~n}a, Mariana and Parra De Freitas, H{\'e}ctor and Sethi, Savdeep",
    title = "{Non-supersymmetric heterotic strings on a circle}",
    eprint = "2307.13745",
    archivePrefix = "arXiv",
    primaryClass = "hep-th",
    doi = "10.1007/JHEP12(2024)082",
    journal = "JHEP",
    volume = "12",
    pages = "082",
    year = "2024"
}

@article{Larotonda:2024thv,
    author = "Larotonda, Vittorio and Lin, Ling",
    title = "{Anomaly inflow and gauge group topology in the 10d Sugimoto string theory}",
    eprint = "2412.17894",
    archivePrefix = "arXiv",
    primaryClass = "hep-th",
    doi = "10.1007/JHEP06(2025)136",
    journal = "JHEP",
    volume = "06",
    pages = "136",
    year = "2025"
}

@article{Fraiman:2026ltu,
    author = "Fraiman, Bernardo and Larotonda, Vittorio and Lin, Ling and Raucci, Salvatore",
    title = "{Non-supersymmetric dualities beyond the gauge algebra}",
    eprint = "2606.30712",
    archivePrefix = "arXiv",
    primaryClass = "hep-th",
    month = "6",
    year = "2026"
}

@article{Polchinski:1995df,
    author = "Polchinski, Joseph and Witten, Edward",
    title = "{Evidence for heterotic - type I string duality}",
    eprint = "hep-th/9510169",
    archivePrefix = "arXiv",
    reportNumber = "IASSNS-HEP-95-81, NSF-ITP-95-135",
    doi = "10.1016/0550-3213(95)00614-1",
    journal = "Nucl. Phys. B",
    volume = "460",
    pages = "525--540",
    year = "1996"
}

@article{Buratti:2021fiv,
    author = "Buratti, Ginevra and Calder{\'o}n-Infante, Jos{\'e} and Delgado, Matilda and Uranga, Angel M.",
    title = "{Dynamical Cobordism and Swampland Distance Conjectures}",
    eprint = "2107.09098",
    archivePrefix = "arXiv",
    primaryClass = "hep-th",
    doi = "10.1007/JHEP10(2021)037",
    journal = "JHEP",
    volume = "10",
    pages = "037",
    year = "2021"
}

@article{Angius:2022aeq,
    author = "Angius, Roberta and Calder{\'o}n-Infante, Jos{\'e} and Delgado, Matilda and Huertas, Jes{\'u}s and Uranga, Angel M.",
    title = "{At the end of the world: Local Dynamical Cobordism}",
    eprint = "2203.11240",
    archivePrefix = "arXiv",
    primaryClass = "hep-th",
    reportNumber = "IFT-UAM/CSIC-22-31",
    doi = "10.1007/JHEP06(2022)142",
    journal = "JHEP",
    volume = "06",
    pages = "142",
    year = "2022"
}

@article{Calderon-Infante:2026ymy,
    author = "Calder{\'o}n-Infante, Jos{\'e} and Cheng, Gongrui and Herr{\'a}ez, Alvaro and Van Riet, Thomas",
    title = "{End-of-the-World Singularities: The Good, the Bad, and the Heated-up}",
    eprint = "2603.18133",
    archivePrefix = "arXiv",
    primaryClass = "hep-th",
    month = "3",
    year = "2026"
}

@article{Makridou:2026jzy,
    author = "Makridou, Andriana and G{\'o}mez, Alejandro Javier Puga",
    title = "{Sharpened Dynamical Cobordism}",
    eprint = "2605.06793",
    archivePrefix = "arXiv",
    primaryClass = "hep-th",
    reportNumber = "IFT-UAM/CSIC-26-45",
    month = "5",
    year = "2026"
}

\end{document}